\documentclass[preprint,12pt]{elsarticle}

\usepackage{amssymb}
\usepackage{xcolor}
\AtBeginDocument{\pagenumbering{arabic}}
\usepackage{stackrel}
\usepackage{amssymb}
\usepackage{amsmath}
\usepackage{multirow}
\usepackage{mathtools}

\usepackage[utf8]{inputenc}
\usepackage[T1]{fontenc}

\usepackage{hyperref}
\usepackage{nomencl}
\usepackage{etoolbox}
\makenomenclature
\renewcommand\nomgroup[1]{%
  \item[\bfseries
  \ifstrequal{#1}{G}{Greek Symbols}{
  \ifstrequal{#1}{S}{Subscripts}{
  \ifstrequal{#1}{P}{Superscripts}{}}}
]}

\journal{Electrochimica Acta}

\newlength{\margin}
\newcommand{\VII}{{\rm V}^{\sf II}}
\newcommand{\VIII}{{\rm V}^{\sf III}}
\newcommand{\VIV}{{\rm V}^{\sf IV}}
\newcommand{\VV}{{\rm V}^{\sf V}}

\newcommand{\CII}{C_2}
\newcommand{\CIII}{C_3}
\newcommand{\CIV}{C_4}
\newcommand{\CV}{C_5}

\newcommand{\XII}{X_2}
\newcommand{\XIII}{X_3}
\newcommand{\XIV}{X_4}
\newcommand{\XV}{X_5}

\newcommand{\RMSE}{E}

\begin{document}

\begin{frontmatter}

%% Title, authors and addresses

%% use the tnoteref command within \title for footnotes;
%% use the tnotetext command for theassociated footnote;
%% use the fnref command within \author or \address for footnotes;
%% use the fntext command for theassociated footnote;
%% use the corref command within \author for corresponding author footnotes;
%% use the cortext command for theassociated footnote;
%% use the ead command for the email address,
%% and the form \ead[url] for the home page:
%% \title{Title\tnoteref{label1}}
%% \tnotetext[label1]{}
%% \author{Name\corref{cor1}\fnref{label2}}
%% \ead{email address}
%% \ead[url]{home page}
%% \fntext[label2]{}
%% \cortext[cor1]{}
%% \affiliation{organization={},
%%             addressline={},
%%             city={},
%%             postcode={},
%%             state={},
%%             country={}}
%% \fntext[label3]{}

\title{A comprehensive guide for measuring total vanadium concentration and state of charge of vanadium electrolytes using UV-Visible spectroscopy}

%% use optional labels to link authors explicitly to addresses:
%% \author[label1,label2]{}
%% \affiliation[label1]{organization={},
%%             addressline={},
%%             city={},
%%             postcode={},
%%             state={},
%%             country={}}
%%
%% \affiliation[label2]{organization={},
%%             addressline={},
%%             city={},
%%             postcode={},
%%             state={},
%%             country={}}

\author[aff1]{Ange A. Maurice\corref{corr1}}
\ead{amaurice@uc3m.es}
\author[aff1,aff2]{Alberto E. Quintero}

\author[aff1]{Marcos Vera}
\ead{marcos.vera@uc3m.es}
\affiliation[aff1]{organization={Departamento de Ingeniería Térmica y de Fluidos, Universidad Carlos III de Madrid},%Department and Organization
            addressline={Avd. de la Universidad 30}, 
            city={Leganés},
            postcode={28911}, 
            state={Madrid},
            country={Spain}}

\affiliation[aff2]{organization={R\&D Department, Micro Electrochemical Technologies},%Department and Organization
            city={Leganés},
            postcode={28918}, 
            state={Madrid},
            country={Spain}}

\cortext[corr1]{Corresponding author}

\begin{abstract}
This paper presents an exhaustive how-to guide on measuring the total vanadium concentration and state of charge of vanadium electrolytes using UV-Visible spectroscopy. The study is provided with an open-access database (\href{https://github.com/AngeAM/SOC_Vanadium_Spectra_2023.git}{GitHub}) that supports the methods and procedures and facilitates access to the calibration data. The study covers the three types of electrolyte solutions relevant to vanadium redox flow batteries, namely the anolyte $\VII/\VIII$, the catholyte $\VIV/\VV$, and the $\VIII/\VIV$ commercial electrolyte, meant to be preconditioned to either $\VIII$ or $\VIV$ before battery operation. Analytical expressions to calculate the concentration of different vanadium species in the electrolyte solutions are provided based on either empirical correlations or spectral deconvolution methods.
The paper also examines the limitations of the measurement technique and provides insightful recommendations for future research. The open-access database provided by the authors is expected to serve as a valuable repository for scholars and scientists working in the field of vanadium redox flow batteries.
\end{abstract}

%%Research highlights
\begin{highlights}
\item UV-Visible spectroscopy is used to characterize vanadium electrolytes.
\item Three electrolyte solutions relevant to vanadium redox flow batteries are studied: the anolyte $\VII/\VIII$, the catholyte  $\VIV/\VV$, and the intermediary electrolyte $\VIII/\VIV$.
\item A fast empirical method and a method based on spectral deconvolution are proposed for each electrolyte.
\item The proposed methods are calibrated over a wide range of concentrations (0.91-1.83M).
\item Total vanadium concentration and  state of charge can be accurately measured.
\item Overall accuracy is $25$-$35$~mM for the concentration and $1.0$-$1.5 \%$ for the state of charge.

\end{highlights}

\begin{keyword}
%% keywords here, in the form: keyword \sep keyword
UV-Vis spectroscopy \sep Vanadium electrolytes \sep Calibration methods \sep Concentration \sep State of charge \sep Redox flow batteries
%% PACS codes here, in the form: \PACS code \sep code

%% MSC codes here, in the form: \MSC code \sep code
%\MSC 76E06 \sep 76R05 \sep 76R10
%% or \MSC[2008] code \sep code (2000 is the default)

\end{keyword}

\end{frontmatter}

%% \linenumbers

% Article body
%\tableofcontents % Auxiliar para ver la estructura rapidamente

% Main text
\section{Introduction}\label{section_intro}

The rapid growth of renewable energy sources as a sustainable alternative to traditional power generation requires the development of effective energy storage solutions capable of mitigating the power grid fluctuations inherent to clean energy technologies \cite{zhao2020renewable}. In this context, vanadium redox flow batteries (VRFBs) offer several advantages that make them a promising large-scale stationary energy storage solution: high round-trip energy efficiency, excellent scalability, long cycle life, decoupled power and energy capacity, deep discharge capability, and high safety compared to other types of batteries. These advantages posit VRFBs as a promising energy storage technology for various applications, including grid-level energy storage, renewable energy integration, load balancing, and backup power systems. However, VRFBs also face challenges such as lower energy and power density, high cost, complex system design, electrolyte degradation, and environmental concerns \cite{cunha2015vanadium, yuan2019review, lourenssen2019vanadium}. 

In particular, one of the main drawbacks of VRFBs is their capacity decay during cycling, mainly caused by ion crossover \cite{yang2015effects}, hydrogen evolution \cite{wei2017insitu}, or imperfect electrolyte mixing in the tanks \cite{prieto2023fluid}. This makes the total vanadium concentration and the state of charge (SOC) to drift away from symmetry between the negative and positive sides, creating a charge imbalance. To address and correct these issues effectively, real-time monitoring of the total vanadium concentration and SOC of the positive and negative half-cells would be highly desirable. Upon charge and discharge, the absorbance spectra of vanadium compounds vary significantly. Thus, a direct, non-invasive method to measure the SOC is through Ultraviolet-Visible (UV-Vis) absorbance spectroscopy. 

The method comes as a valuable alternative to the classical approach of open circuit voltage (OCV) measurements, typically conducted offline in a smaller electrochemical cell separate from the main redox reactor. These OCV sensors are considered invasive due to the potential for crossover and self-discharge through the membrane of the smaller cell. Moreover, they are susceptible to temperature variations, and their measurement accuracy heavily depends on the precise design and operation of the cell to mitigate issues such as mass transfer limitations, preferential paths, or shunt currents, among others. Moreover, OCV measurements solely provide information about the overall SOC of the VRFB system, lacking the ability to determine the individual SOC of each electrolyte (positive and negative). Additionally, they do not provide insights into the total Vanadium concentration in each stream~\cite{knehr2011open,munoz2023exploring}.

More recent approaches, such as the online monitoring of density or viscosity to establish their correlation with the state of charge (SOC), still have limitations, including temperature dependency and the introduction of an additional pressure drop. These factors contribute to a decrease in the overall round-trip efficiency of VRFBs~\cite{li2018investigation, ressel2018state}. In the case of membraneless redox flow batteries, which have gained significant research attention, an optical non-invasive measurement becomes crucial due to the limitations associated with the aforementioned methods. Furthermore, in membraneless systems, the simultaneous measurement of SOC and the total vanadium concentration in both electrolytes is required due to the absence of a membrane.
Specifically, for micro membraneless VRFBs, it is imperative to minimize the dead volume in the microsensor, a challenge effectively addressed through non-invasive optical methods~\cite{navalpotro2023neutral,lee2013microfluidic}. Overall, optical sensors enable monitoring VRFB performance and provide more profound insights into the specific processes that control efficiency and capacity decay during operation, offering optimal real-time readings and enabling closed-loop control strategies~\cite{di2022general,monbaliu2023will}.

In the literature, the UV-Vis spectroscopy method has been extensively studied~\cite{geiser_photometrical_2019-1, roznyatovskaya_detection_2016, brooker_determining_2015, choi_analysis_2013}, uncovering early the complex chemistry of the positive electrolyte with Blanc et al.~\cite{blanc_spectrophotometric_1982} and more recently with Buckley et al.~\cite{buckley_towards_2014, quill_factors_2015, gao_spectroscopic_2013, petchsingh_spectroscopic_2016}. Experimental tests have also been conducted using the UV-Vis method to determine the state of charge (SOC) both offline~\cite{liu_state_2012, petchsingh_spectroscopic_2016} and online/in-operando~\cite{tang_monitoring_2012, zhang_-line_2015, shin_real-time_2020}. However, upon browsing the extensive literature on the subject, a critical question still emerges: How can the total vanadium concentration and the SOC of both electrolytes in a VRFB be independently measured using UV-Vis spectroscopy, and what level of accuracy can be achieved with this method? While partial answers can be found in scattered publications, the available information often focuses solely on one specific electrolyte and lacks detailed analysis. In addition, certain reported calibrations only provide the SOC as an output, disregarding the total vanadium concentration. Moreover, limited information is available regarding measurement errors, making it difficult to evaluate the robustness of the reported calibration methods. Furthermore, existing studies often present calibrations for a narrow range of concentrations.

To effectively address these knowledge gaps, the primary objective of this article is to present a comprehensive document that outlines calibration methods capable of accurately quantifying both the total vanadium concentration and the SOC across all possible vanadium electrolyte mixtures, encompassing a wide concentration range of 0.91 to 1.83~M. Moreover, we provide significant calibration improvements compared with the existing literature, resulting in better accessibility for the reader and improved measurement accuracy.

This study deals with the three types of vanadium electrolyte mixtures employed in VRFBs: \textit{i)} the anolyte $\rm V^{2+}/V^{3+}$ ($\VII/\VIII$), \textit{ii)} the $\rm V^{3+}/VO_2^{+}$ ($\VIII/\VIV$) mixture and \textit{iii)} the catholyte $\rm VO^{2+}/VO_2^{+}$ ($\VIV/\VV$). Following standard practice, the four oxidation states of vanadium are denoted here $\VII$, $\VIII$, $\VIV$, and $\VV$. During normal VRFB operation, the $\VIII/\VIV$ electrolyte should not be present and is typically less relevant. Nonetheless, a failing VRFB with high electrolyte imbalance could overdischarge and result in $\VIII/\VIV$ appearing in one of the tanks. Moreover, commercial vanadium electrolyte is usually distributed as an equimolar $\VIII/\VIV$ mixture, often referred to as $\rm V{3.5+}$, meant to undergo a preconditioning process to achieve zero SOC either as anolyte or catholyte. Other vanadium electrolyte mixtures cannot exist since self-discharge reactions would occur yielding one of the three composition pairs indicated above~\cite{Tang_thermal_2012}.

For each electrolyte we provide two calibration methods: \textit{i)} a fast empirical method based on linear regression, which requires only one or two hand-picked absorbances as input, and \textit{ii)} a more computationally intensive method based on spectral deconvolution~\cite{loktionov_operando_2022}, which uses the whole spectrum as input. The methods to be outlined below are based on the generalized Beer-Lambert law. This law states that the total absorbance of a mixture containing multiple absorbing compounds is equal to the sum of the individual absorbances contributed by each compound. This principle assumes that the absorbing compounds in the mixture do not interact with each other and that their absorbance contributions are independent~\cite{atkins2014physical, harris2015quantitative, skoog2013fundamentals}. 

The calibration methods yield the total vanadium concentration $C$ and the mole fractions $X_i$ for the three electrolyte mixtures under study. In the case of the $\VII/\VIII$ and $\VIV/\VV$ mixtures, anolyte and catholyte, the calibration also provides the SOC, represented by the mole fractions $\XII$ and $\XV$, which is the desired piece of information for most readers. Table \ref{tab:notation} defines the notation used in this paper, relating the total vanadium concentration $C$, the mole fractions $X_i$, and the SOC to the molar concentrations $C_i$ of the four vanadium ions, $i = \{2, 3, 4, 5\}$. Note that in the $\VIV/\VV$ mixture we include the presence of the 1:1 stoichiometric mixed-valence complex $\rm V_2O_3^{3+}$ in equilibrium with $\VIV$ and $\VV$ according to reaction \eqref{eq:V2O33p_formation}, whose concentration is denoted by subscript $i = 45$ \cite{blanc_spectrophotometric_1982}. Hereafter, the SOC and mole fractions are expressed as percentages unless otherwise stated. 

The data set and the Python calibration algorithms generated in this work are publicly shared in an online repository (\href{https://github.com/AngeAM/SOC_Vanadium_Spectra_2023.git}{GitHub}). We believe this decision will be helpful for both the scientific community and the industry. In general, adopting an open data approach proves most valuable in advancing the field, as it allows anyone to enhance the existing calibration procedures by employing more advanced chemometric techniques.

\begin{table}[b!]
    \centering
    \begin{tabular}{cccc}
        & \multicolumn{3}{c}{Electrolyte mixture} \\[1mm] \cline{2-4} \\[-3.5mm]
        Variable & $\VII/\VIII$ & $\VIII/\VIV$ & $\VIV/\VV$\\[1mm] 
        \hline \\[-3.5mm]
        $C$     & $\CII+\CIII$ & $\CIII+\CIV$ & $\CIV+\CV+2C_{45}$ \\[1mm] 
        \hline \\[-3.5mm]
        $\XII$  & $\dfrac{\CII}{\CII+\CIII}$ & -- & -- \\[4mm]
        $\XIII$ & $\dfrac{\CIII}{\CII+\CIII}$ & $\dfrac{\CIII}{\CIII+\CIV}$ & -- \\[4mm]
        $\XIV$  & -- & $\dfrac{\CIV}{\CIII+\CIV}$ & $\dfrac{\CIV+C_{45}}{\CIV+\CV+2C_{45}}$ \\[4mm]
        $\XV$   & -- & -- & $\dfrac{\CV+C_{45}}{\CIV+\CV+2C_{45}}$ \\[4mm] 
        \hline \\[-3.5mm]
        SOC     & $X_2$ & -- & $X_5 = 1-X_4$
    \end{tabular}
    \caption{Definition of the total vanadium concentration $C$, molar fractions $X_i$, and SOC for the different electrolyte mixtures under study as a function of the molar concentration of the four vanadium ions $C_i$, $i = \{2, 3, 4, 5\}$. In the $\VIV/\VV$ mixture $C_{45}$ denotes the concentration of the 1:1 stoichiometric mixed-valence complex $\rm V_2O_3^{3+}$ at equilibrium with $\VIV$ and $\VV$ according to reaction \eqref{eq:V2O33p_formation}.}
    \label{tab:notation}
\end{table}

The paper is organized as follows. Section \ref{sec:experimental} describes the experimental procedures employed to prepare the calibration samples with the best accuracy. Section \ref{sec:results_and_calibration} presents the calibration methods for the different mixtures, starting with the anolyte $\VII/\VIII$ and the $\VIII/\VIV$ mixture, which exhibit a common linear response. Subsequently, the calibration of the catholyte $\VIV/\VV$ is discussed, showcasing its non-linear spectral response. In Section \ref{sec:discussion}, we discuss the accuracy, advantages and disadvantages of the different methods. Finally, in Section \ref{sec:conclusions} we summarize the key findings and draws the conclusions.

\section{Experimental}\label{sec:experimental}

\subsection{Electrochemical preparation of the reference electrolytes} \label{sec:prep_chem}

The calibration of UV-Vis spectroscopic methods requires the preparation of single-species reference electrolytes, and of mixtures thereof, with a high degree of compositional precision. In this section we describe the experimental procedures employed to generate these samples. We prepared the reference solutions of vanadium electrolytes via electrochemical reduction and oxidation in an in-house 3 $\rm cm^2$ electrochemical cell. To avoid hydrogen evolution reactions and significant crossover effects, the applied cell voltage was kept at 1.6~V. Throughout the charge process, we acquired UV-Vis absorbance spectra periodically using flow cuvettes. As the starting electrolyte, we used an equimolar $\VIII/\VIV$ commercial solution (Oxkem Limited, UK) containing an average oxidation number (AOS) of $+3.5$. Its exact composition is listed in Table \ref{tab:fresh_composition}. Since all samples originated from this batch solution, we performed induction coupled plasma-optical emission spectroscopy (ICP-OES) to measure the total vanadium concentration with high accuracy, with the results given in the table. It is interesting to note that the solution contains 0.5~M of phosphoric acid, which is commonly used in commercial vanadium electrolytes to improve their solubility properties~\cite{oldenburg_revealing_2018}.
\begin{table}[ht!]
    \centering
    \begin{tabular}{cc}
         Species & Concentration \\[1mm]
         \hline \\[-3.5mm]
        Vanadium & $1.83 \pm 0.092$~M \\
        Sulfuric Acid & 4.6~M \\
        Phosphoric Acid & 0.5~M \\[1mm]
         \hline \\[-3.5mm]
        Oxidation number & $\approx 3.5$\\
    \end{tabular}
    \caption{Composition of the commercial vanadium electrolyte (Oxkem Limited, UK) used in the study, with an average oxidation number (AOS) of $+3.5$.}
    \label{tab:fresh_composition}
\end{table}

Slight variations of the total vanadium concentration can significantly affect the absorbance spectra measurements. This can occur, for instance, due to ion crossover across the membrane. To minimize such phenomenon, we stacked 3 layers of Nafion\textsuperscript{\tiny\textregistered}~N112 membrane following the results published by Ashraf et al.~\cite{ashraf_gandomi_direct_2020}. With this membrane stacking and taking into account our charge duration of 5-15 hours, the total vanadium concentration should not vary more than 5 mM, which represents a variation of about 0.27\%. Note that the calibrations and data shown in this paper are provided for a given acidity level. Minimal variations of spectrum shape could appear when varying the acidity~\cite{loktionov_operando_2022}. However, given the small magnitude of these variations, we expect negligible impacts on the accuracy of the calibration. Additionally, the methodology presented in this paper remains valid regardless of the acidity and could be used to update the calibration for various acidic conditions, a potential research work that is outside the scope of this paper.

\subsection{UV-Vis spectroscopy}\label{sec:UV_vis}

We acquired the spectra using 0.1~mm and 1~mm path length commercial flow cuvettes (Starna Scientific Ltd., UK). Because of its lower absorption, we measured the $\VII$/$\VIII$ electrolyte using the 1~mm path length cuvette whilst for the other mixtures we used the 0.1~mm path length cuvette. The spectrometer (Flame T-VIS-NIR-ES, Ocean Insight, USA), the lamp (HL-2000-FHSA-LL, Ocean Insight, USA), and the cuvette holder (SQUARE ONE, Ocean Insight, USA) were connected via optical fibers. The Flame T-VIS-NIR-ES spectrometer has a wavelength detection range of 345-1033 nm and an optical resolution of 1.33 nm FWHM. However, considering the high absorbance of the electrolyte and the lamp emission range, the actual measurable range is about 420-1000nm depending on the concentration and the species. Upon measurement, we injected at least 1 mL of sample with a syringe in the flow cuvette in order to ensure ample clearing of the cuvette optical window. In this work, all absorbances are normalized with the optical path length, and thus their units are in $\rm cm^{-1}$.

\subsubsection{Preparation of $\VII$ and $\VV$}

The preparation of the charged $\VII$ and $\VV$ electrolyte solutions was achieved by starting from the equimolar 
$\rm V{3.5+}$ solution and bringing the electrochemical oxidation and reduction of the electrolyte to completion. In other words, fully oxidizing or reducing the vanadium applying a fixed 1.6~V potential to the cell. The completion point was deemed to be reached when three independent criteria were met: \textit{i)} the current density was lower than $2~\rm mA/cm^2$, \textit{ii)} the UV-Vis spectrum remained stable for more than 10 min, and \textit{iii)} the shape of the spectrum was coherent with previous literature results. It is worth noting that we had to increase the applied potential up to 1.7-1.8~V for a short period of time (10-15~min) to achieve a stable UV-Vis spectrum.

\subsubsection{Preparation of $\VIII$ and $\VIV$}

Similarly, we prepared the discharged $\VIII$ and $\VIV$ electrolytes by electrochemical charging the equimolar $\rm V{3.5+}$ solution. However, reaching the exact transition state, or endpoint, where only pure $\VIII$ or $\VIV$ exist is more complex than for the fully charged electrolytes $\VII$ and $\VV$. In this case, the use of online spectrometry is critical. The procedure, called spectrophotometric titration \cite{skoog2013fundamentals}, consists in lowering the applied voltage or current when approaching the purity point and monitoring the spectra at specific wavelengths looking for an inflection point. Further elaboration on the procedure and specific criteria for its application to vanadium electrolytes can be found in the Supplementary Information (Figure~S1). Obtaining $\VIII$ is relatively easier through this method, as an overcharge leads to the formation of $\VII$, which subsequently undergoes air oxidation. Indeed, when the negative electrolyte is exposed to air, it naturally tends to become pure $\VIII$. Nevertheless, in order to prevent any potential impact on the total vanadium concentration due to water evaporation, efforts were made to minimize exposure to air.

\subsubsection{Calibration samples}

We prepared the calibration samples via pipetting and mixing/diluting of the reference solutions and a 4.6M sulfuric acid solution. The total sample volumes were 1-2~mL, depending on the desired vanadium concentration. Using deionized water and a 5-digit analytical weighing scale, we estimated the pipetting accuracy to be 0.4\%. For each mixture, we prepared 11 samples at SOCs ranging from 0 to 100\% and concentrations of 1.83, 1.52, 1.22, and 0.91~M (44 samples per mixture, 132 samples in total). We disregarded concentrations lower than 0.9~M for the calibration step since most industrial applications use electrolytes with concentrations higher than 1~M to maintain high capacity/volume ratios. Nonetheless, to verify the Beer-Lambert law, we additionally prepared samples with 0.6 and 0.3~M containing only pure compounds (see Section~\ref{sec:ref_spectra}).

\section{Results and calibration}
\label{sec:results_and_calibration}

\subsection{Reference spectra}
\label{sec:ref_spectra}
\begin{figure}[ht!]
    \centering
    \includegraphics[scale=0.525]{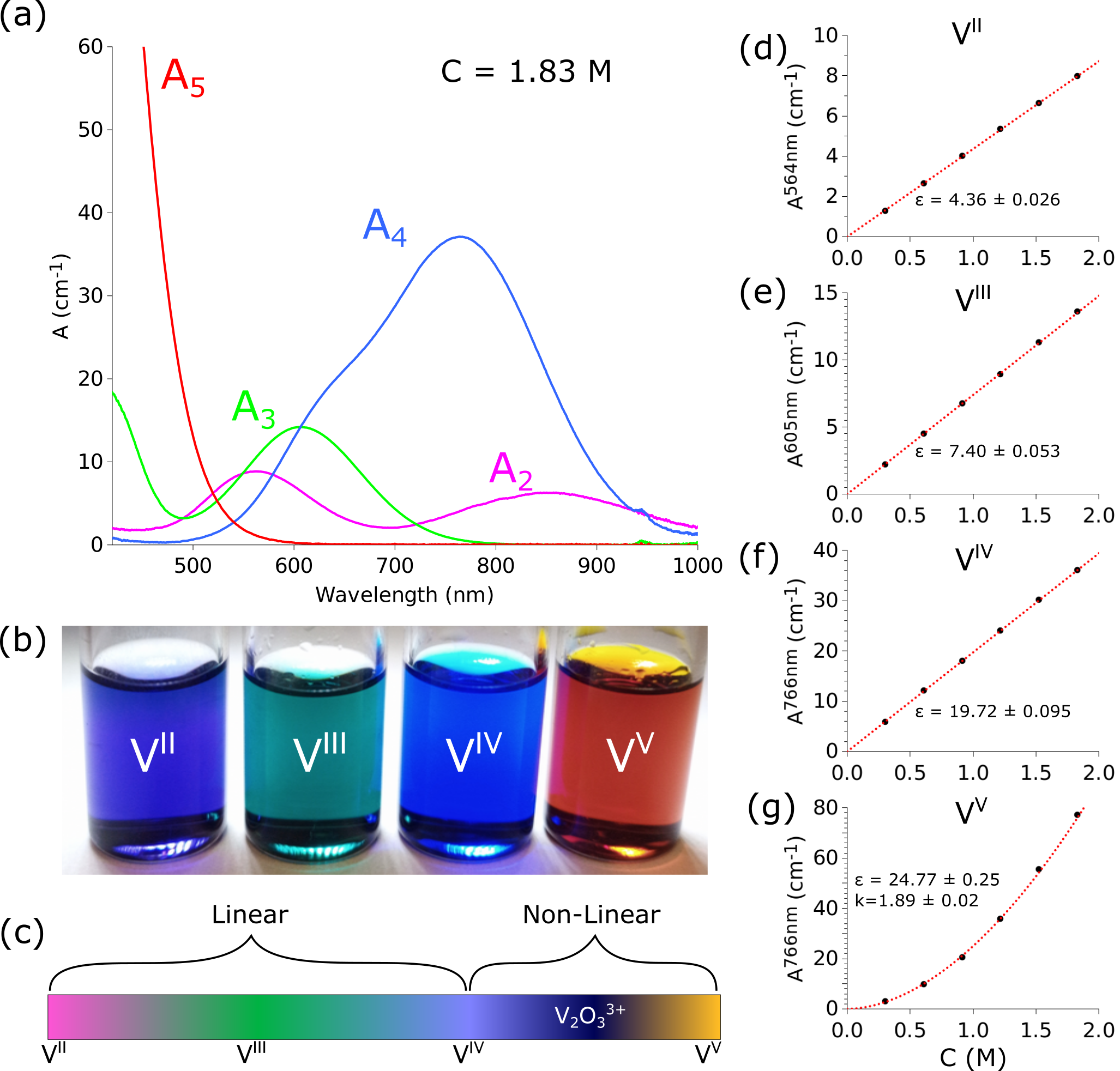}
    \caption{(a) 1.83~M reference absorbance spectra of the electrolyte solutions containing the four pure vanadium oxidation states. (b) Picture of the vials containing solutions of the pure vanadium electrolytes. The solutions were diluted when needed for visual purposes. (c) Color band representing the visual appearance of the vanadium electrolytes. The brackets indicate the locations where the spectral response is linear and non-linear. (d, e, f, g) Absorbance of $\VII$, $\VIII$, $\VIV$ and $\VV$ extracted, respectively at 564, 605, 766, and 440~nm as a function of the concentration. The experimental points are plotted as black circles while the fit is plotted as a red dotted line.}
    \label{fig:ref_spectra}
\end{figure}

Figure \ref{fig:ref_spectra}a shows the reference absorbance spectra $A_i$ of the single-species electrolyte solutions, obtained following the procedures outlined in section~\ref{sec:experimental}. The spectra $A_2$, $A_3$, $A_4$, and $A_5$ correspond, respectively, to the oxidation states $\VII$, $\VIII$, $\VIV$, and $\VV$. The shapes of the spectra are consistent with the visual appearance of the electrolytes depicted in Figure \ref{fig:ref_spectra}b, as well as the color band shown in Figure \ref{fig:ref_spectra}c. For instance, $A_5$ acts as a high-pass filter, suppressing the blue part of the visible spectrum and having negligible absorbance above 650 nm, which yields an orange/yellow color. By contrast, $A_4$ has very low absorption between 420-550 nm, resulting in a deep blue appearance. Figures \ref{fig:ref_spectra}d, \ref{fig:ref_spectra}e, \ref{fig:ref_spectra}f, and \ref{fig:ref_spectra}g show the measured absorbance as a function of the total vanadium concentration for the four vanadium oxidation states. The plots correspond to selected wavelengths, approximately aligned with the absorbance peaks depicted in Figure \ref{fig:ref_spectra}b for $\VII$, $\VIII$, and $\VIV$, and at 440 nm for $\VV$, which lacks a well defined peak in the observed range. The linear fits (red dotted lines) obtained for $\VII$, $\VIII$, and $\VIV$ confirm the validity of the Beer-Lambert law 
\begin{equation}
\label{eq:beer_lambert}
    A_i = \epsilon_i C_i \quad \text{for} \quad i=\{2, 3, 4\}
\end{equation}
at least up to total vanadium concentrations of 1.83~M. In this expression $A_i$ is the path-length normalized absorbance spectrum in $\rm cm^{-1}$, $C_i$ is the molar concentration in M, and $\epsilon_i$ represents the molar absorptivity (or molar absorption coefficient) of the $i$-th oxidation state, in $\rm cm^{-1}~M^{-1}$.

By contrast, the relation between absorbance and concentration for $\VV$ is non-linear, deviating from the classical Beer-Lambert law. This non-linear behavior can be effectively described by a power law equation of the form
\begin{equation}
\label{eq:beer_lambert_power}
A_5 = \epsilon_5 C_5^k
\end{equation}
where the fitting exponent remains approximately constant for different wavelengths with a value of roughly $k \approx 2$, and $\epsilon_5$ represents the molar absorptivity of $\VV$. 
The non-linear absorbance of $\VV$ had been only briefly reported for $\VV$ at ultra-low concentrations ($<0.03$~M) by Bannard et al.~\cite{bannard_spectrophotometric_1968}. Consequently, the deviation observed here at high concentrations is a novel result that, to the best of our knowledge, had not been previously reported in the literature. Non-linear deviations from the Beer-Lambert law at high concentrations may be due to different factors, such as intermolecular interactions or changes in the refractive index of the analyte. These deviations may result in complex spectral behaviors that cannot be accurately described by a simple linear relationship between absorbance and concentration \cite{mayerhofer2020bouguer, nolte2021IR, shin_real-time_2020}.

The values of $\epsilon_i$ and $k$ for the chosen wavelengths are annotated in the graphs of Figure \ref{fig:ref_spectra} along with their corresponding statistical uncertainties given by the root mean square error (RMSE, or $E$ in shorthand notation) between the predicted and the measured values. 
Table \ref{tab:val_ref_epsilon} lists values of the molar absorptivities $\epsilon_i$ for $i=\{2,3,4\}$ at selected wavelengths. As previously discussed, these wavelengths correspond to the local absorbance peaks of $\VII$, $\VIII$, and $\VIV$, with the corresponding maxima of $\epsilon_i$ being highlighted in boldface in each case.

\begin{table}[ht!]
    \centering
    \begin{tabular}{cccc}
        & \multicolumn{3}{c}{Electrolyte} \\[3mm]
        & $\VII$ & $\VIII$ & $\VIV$ \\[1mm] 
        \cline{2-4} \\[-3.5mm]
        Wavelength&$\epsilon_2$ & $\epsilon_3$ & $\epsilon_4$ \\[1mm]
        \hline \\[-3.5mm]
        564 nm & $\mathbf{4.36} \pm 0.026$ & $5.46 \pm 0.079$          & $2.91 \pm 0.043$ \\
        605 nm & $3.21 \pm 0.020$          & $\textbf{7.40}\pm 0.053$ & $7.18 \pm 0.061$ \\
        766 nm & $2.11 \pm 0.014$          & negligible          & $\textbf{19.72} \pm 0.095$ \\
        850 nm & $\textbf{3.18} \pm  0.032$ & negligible          & $10.82 \pm 0.19$
    \end{tabular}
    \caption{Values of the 1.83~M molar absorptivities $\epsilon_i$ for $i=\{2,3,4\}$ at selected wavelengths determined by the local maxima of the absorbance curves of $\VII$, $\VIII$, and $\VIV$. All units are in $\rm cm^{-1} M^{-1}$.}
    \label{tab:val_ref_epsilon}
\end{table}

The following sections provide detailed information about the spectral responses of the mixtures $\VII/\VIII$, $\VIII/\VIV$, and $\VIV/\VV$. While the $\VII/\VIII$ and $\VIII/\VIV$ mixtures show a linear response, the catholyte $\VIV/\VV$ exhibits a non-linear response. This non-linear behavior is attributed both to the non-linear absorbance of $\VV$ and to the formation of a high absorbance 1:1 stoichiometric complex between $\VIV$ and $\VV$~\cite{blanc_spectrophotometric_1982}. Consequently, the calibration methods for the three electrolyte mixtures are presented in two distinct sections addressing the linear and non-linear cases separately.

\subsection{Linear case: $\VII/\VIII$ and $\VIII/\VIV$ }

\begin{figure}[b!]
    \centering
    \includegraphics[scale=0.63]{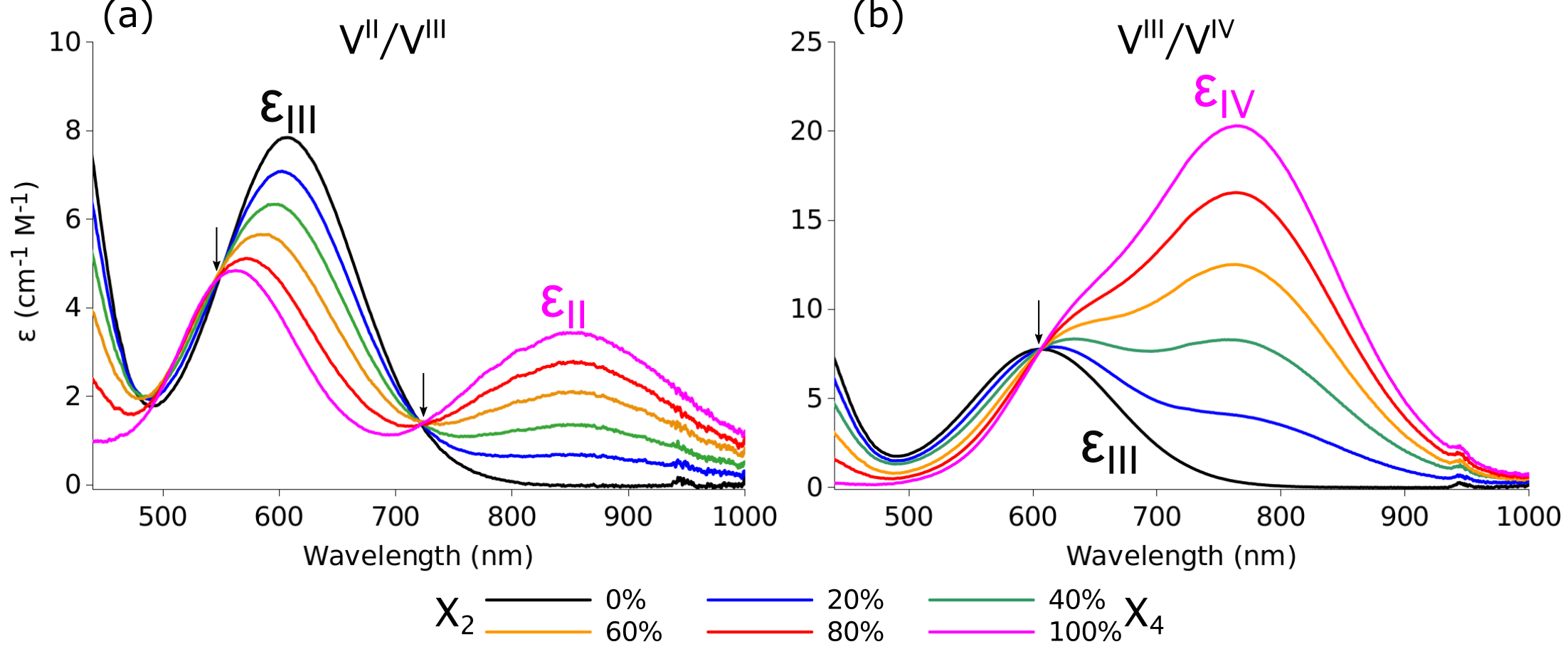}
    \caption{Molar absorptivity spectra of the $\VII/\VIII$ and $\VIII/\VIV$ at varying SOC/mole fractions for a total vanadium concentration $C = 1.83$~M. The arrows indicate the location of the isosbestic points. The percentage in the legend represents the mole fractions of (a) $\VII$, which coincides with the SOC, and (b) $\VIV$, so that the black line corresponds to the molar absorption coefficient of $\VIII$ in both subplots.}
    \label{fig:spectra_soc}
\end{figure}

Figures \ref{fig:spectra_soc}a and \ref{fig:spectra_soc}b show, respectively, the molar absorptivity spectra of the $\VII/\VIII$ and $\VIII/\VIV$ mixtures at varying SOCs, or mole fractions, for a total vanadium concentration $C = 1.83$~M. The former electrolyte is characterized by the mole fraction of $\VII$, or $X_2$, which coincides with the SOC, while the later is characterized by the mole fraction of $\VIV$, or $X_4$. The vertical arrows indicate the location of the isosbestic points in which the absorbance remains constant regardless of the mole fraction. In UV-Vis spectroscopy, an isosbestic point refers to a specific wavelength at which the absorbance of the sample remains constant during a chemical reaction or a change in the sample's composition. 
The presence of such points suggests that the reaction occurring in the sample proceeds via a direct conversion between species, without any intermediates, e.g., reaction $\VII \rightleftharpoons \VIII$. 

When dealing with mixtures, the Beer-Lambert law \eqref{eq:beer_lambert} can be extended to account for multiple absorbing species. Therefore, for a binary mixture of two pure compounds A and B, with mole fractions $X_{\rm A}$ and $X_{\rm B} = 1- X_{\rm A}$, the absorption spectrum is the sum of the individual absorbances of the two components
\begin{equation}\label{eq:lin_comb}
    A_{\rm M} = [\epsilon_{\rm A}X_{\rm A} + \epsilon_{\rm B}(1-X_{\rm A})]C
\end{equation}
where $A_{\rm M}$ is the absorbance of the mixture, $\epsilon_{\rm A}$ and $\epsilon_{\rm B}$ are the molar absorptivities of compounds A and B, and $C$ is the total (vanadium) concentration.
At the isosbestic point $\epsilon_{\rm A} = \epsilon_{\rm B} \equiv \epsilon^{\rm iso}$ and Equation~\eqref{eq:lin_comb} reduces to
\begin{equation}\label{eq:iso}
    A_{\rm M}^{\rm iso} = \epsilon^{\rm iso} C
\end{equation}
In other words, at the isosbestic point the absorbance is independent of the mole fraction $X_{\rm A}$ and grows linearly with the total concentration $C$.
Table \ref{tab:val_isosbestic} lists the isosbestic wavelengths, $\lambda^{\rm iso}$, and the corresponding absorbance values, $\epsilon^{\rm iso}$, for the two optically linear vanadium mixtures. As illustrated below, the presence of isosbestic points enables the estimation of the total concentration $C$ independently of the mole fraction $X_{\rm A}$ by directly using the values of $A_{\rm M}^{\rm iso}$ and $\epsilon^{\rm iso}$ in Equation~\eqref{eq:iso}.

The negative electrolyte ($\VII$/$\VIII$) appears to have three isosbestic points. However, only two of them are listed in Table \ref{tab:val_isosbestic} because the third one, located around 500~nm, is too out of focus to have practical application. Equation~\eqref{eq:lin_comb} may lose validity at lower wavelengths ($\lambda<550$) due to uncharted interactions among the vanadium compounds. This hypothesis is consistent with the findings of Loktionov et al.~\cite{loktionov_operando_2022}, whose results suggest the presence of additional interactions at lower wavelengths within the negative electrolyte.

\begin{table}
    \centering
    \begin{tabular}{ccc}
        Electrolyte mixture & $\lambda^{\rm iso}~(\rm nm)$ & $\epsilon^{\rm iso}~(\rm cm^{-1} M^{-1})$ \\[1mm]
      \hline \\[-3.5mm]
      $\VII$/$\VIII$ & $722.11 \pm 0.69 $ & $1.24 \pm 0.001$ \\[1mm]
       & $542.73 \pm 7.03 $ & $4.06 \pm 0.023$ \\[1mm]
      \hline \\[-3.5mm]
      $\VIII$/$\VIV$ & $608.22 \pm 2.33 $ & $7.41 \pm 0.05$ \\ 
    \end{tabular}
    \caption{Wavelengths of the isosbestic points and their corresponding absorbances for the the two optically linear vanadium mixtures.}
    \label{tab:val_isosbestic}
\end{table}

%It is important to note that the presence of sulfuric acid and phosphoric acid in the solutions considered here does not invalidate the assumption of a binary sample underlying Equation \eqref{eq:lin_comb}. Just like water, the main component of the vanadium electrolyte solutions, the absorbances of these acids are several orders of magnitude less than those of the vanadium compounds under study. As a result, their impact on the spectra can be safely disregarded, and the solutions can be effectively treated as binary mixtures for all optical purposes.

\subsubsection{Fast empirical method via linear regression}
\label{sec:empirical234}

According to Equation~\eqref{eq:lin_comb}, by plotting the mole fraction $X_{\rm A}$ versus the absorbance $A_{\rm M}$ at a specific wavelength and performing a linear regression analysis we could obtain a linear relationship that provided $X_{\rm A}$ in terms of $A_{\rm M}$. However, it is important to note that despite the simplicity of this method, it may produce inaccurate results. This is primarily because the absorbance varies with the concentration $C$ and the path length, which may be unknown or altered due to experimental conditions. For example, in the case of VRFBs, species crossover may have a significant impact on the overall concentration of vanadium ions in both the positive and negative electrolytes, thereby affecting the accuracy of the results.

We can address this problem by dividing the total mixture absorbance $A_{\rm M}$ by the absorbance at the isosbestic point $A_{\rm M}^{\rm iso}$, effectively merging Equations~\eqref{eq:lin_comb} and~\eqref{eq:iso} into a single equation
\begin{equation}
\frac{A_{\rm M}}{A_{\rm M}^{\rm iso}} = \frac{(\epsilon_{\rm A} - \epsilon_{\rm B})X_{\rm A} + \epsilon_{\rm B}} {\epsilon^{\rm iso}}
\end{equation}
This equation can be solved explicitly for $X_{\rm A}$ to yield a linear relation of the form
\begin{equation}
X_{\rm A} = a\frac{A_{\rm M}}{A_{\rm M}^{\rm iso}} + b 
\label{eq:SOClin_a_b}
\end{equation}
with $a = \epsilon^{\rm iso}/(\epsilon_{\rm A} - \epsilon_{\rm B})$ and $b = - \epsilon_{\rm B}/(\epsilon_{\rm A}-\epsilon_{\rm B})$. According to this expression, $X_{\rm A}$ should vary linearly with $A_{\rm M}/A_{\rm M}^{\rm iso}$ and be independent of the total concentration $C$. Moreover, the total concentration $C$ could be estimated from Equation~\eqref{eq:iso} as follows
\begin{equation}
\label{eq:totalC_lin}
    C = \frac{A_{\rm M}^{\rm iso}}{\epsilon^{\rm iso}}
\end{equation}
Hence $C$ should grow linearly with the absorbance regardless of the mole fraction $X_{\rm A}$, provided that the absorbance is measured at the isosbestic point.

\begin{figure}[t!]
    \centering
    \includegraphics[scale=0.63]{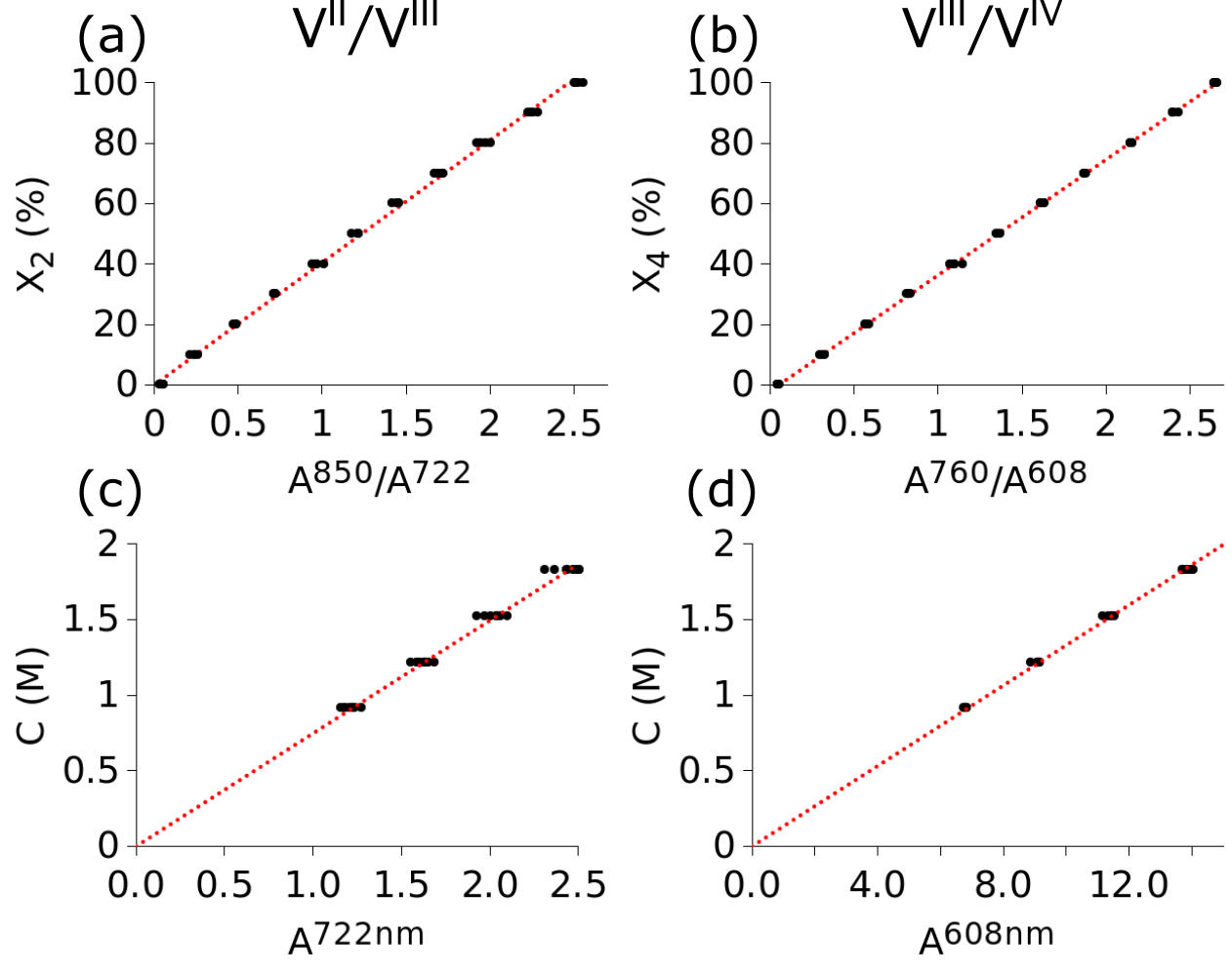}
    \caption{(a, b) Mole fractions $X_2$ and $X_4$ as a function of $A_{\rm M}^\lambda/A_{\rm M}^{\rm iso}$ for the $\VII/\VIII$ ($\lambda = 722$~nm) and $\VIII/\VIV$ ($\lambda = 608$~nm) mixtures. (c, d) Total vanadium concentration as a function of $A_{\rm M}^{\rm iso}$ for the same mixtures. Black dots: experimental values; red dotted lines: linear fits (see Table~\ref{tab:val_lin_reg}). The plots display experimental results from the 44 calibration samples of each mixture, meaning that the linear regressions are performed over a wide range of total vanadium concentrations.}
        \label{fig:lin_reg}
\end{figure}

Figures \ref{fig:lin_reg}a and \ref{fig:lin_reg}b show respectively the mole fractions $X_2$ and $X_4$ of the 44 calibration samples of the $\VII/\VIII$ and $\VIII/\VIV$ mixtures plotted as a function of $A_{\rm M}^\lambda/A_{\rm M}^{\rm iso}$ at a particular wavelength for each mixture, along with the corresponding linear fits \eqref{eq:SOClin_a_b} as detailed in Table~\ref{tab:val_lin_reg}, using the first isosbestic wavelength reported in Table~\ref{tab:val_isosbestic} for each mixture. Similarly, Figures \ref{fig:lin_reg}c and \ref{fig:lin_reg}d show the total concentration of vanadium ions as a function of the isosbestic absorbances $A_{\rm M}^{\rm iso}$ along with the corresponding linear fits \eqref{eq:totalC_lin} also detailed in Table~\ref{tab:val_lin_reg}. The experimental points are shown as black circles, while the linear fits are plotted as red dotted lines. In these plots, all 44 calibration samples are included, meaning that the fits are performed within different vanadium concentrations. The excellent linearity observed in all cases provides strong evidence for the validity and consistency of Equations~\eqref{eq:SOClin_a_b} and \eqref{eq:totalC_lin} across a wide range of total vanadium concentrations and mole fractions. Table~\ref{tab:val_lin_reg} summarizes the correlations thus obtained with their corresponding coefficients calculated from the linear fits, as well as the estimated RMSE, commonly used for reporting the error of calibration methods. 

Thus, when using the linear fits from the current calibration reported in Table~\ref{tab:val_lin_reg} to determine the mole fractions, the SOC, or the total vanadium concentration of a new set of experimental data, the reported values should include, as a minimum, the uncertainty in the measurement by employing the usual notation, e.g., $X = X_{\rm exp} \pm \RMSE_X$.

\begin{table}[ht]
    \centering
\begin{tabular}{ccccc}
    & \multicolumn{4}{c}{Electrolyte mixture} \\[1mm] \cline{2-5} \\[-3.5mm]
    & \multicolumn{2}{c}{$\VII$/$\VIII$} & \multicolumn{2}{c}{$\VIII$/$\VIV$} \\[1mm] \hline \\[-3.5mm]
    Variable & Linear fit & $\RMSE_X$ & Linear fit & $\RMSE_C$ \\[1mm] \hline \\[-3.5mm]
X (\%) & $X_2 = 40.51 \dfrac{A^{850}}{A^{723}}$ & $\pm 1.46$ & $ X_4 = 38.26 \dfrac{A^{760}}{A^{608}} - 1.91$ & $\pm 0.52$  \\
  &  &  &  &  \\
C (M) & $C = \dfrac{1}{1.34} A^{723}$ & $\pm 0.03$ & $C = \dfrac{1}{7.71} A^{608}$  & $\pm 0.015$  \\
 \end{tabular}
\caption{Linear fits for the mole fractions of $\VII$ and $\VIV$ versus $A_{\rm M}/A_{\rm M}^{\rm iso}$ and total vanadium concentration versus $A_{\rm M}^{\rm iso}$ in the $\VII/\VIII$ and $\VIII/\VIV$ mixtures with indication of the corresponding RMSE errors.}
    \label{tab:val_lin_reg}
\end{table}

\subsubsection{Spectral deconvolution}\label{sec:deconv_lin}

The spectral deconvolution method in UV-Visible spectroscopy is a technique used to separate overlapping absorption spectra of multiple components in a mixture. It involves mathematically extracting the individual contributions of each component from the observed composite spectrum, enabling the quantification of their respective concentrations or absorbances. Its application to a mixture of two pure compounds, A and B, starts by expressing the individual absorbances in terms of the known molar absorptivities, $\epsilon_{\rm A}$ and $\epsilon_{\rm B}$, and the unknown molar fractions, $X_{\rm A}$ and $X_{\rm B} = 1 - X_{\rm A}$, and total (vanadium) concentration, $C$. The calibration then proceeds by fitting the best pair ($X_{\rm A}$, $C$) that minimizes the sum of the error squares between the experimentally measured mixture absorbance $A_{\rm exp}$ and the mixture absorbance predicted by the Beer-Lambert law~\eqref{eq:lin_comb}
\begin{equation}
\label{eq:deconv_linear}
\sum_{\lambda} \left\{A_{\rm exp} - \left[\epsilon_{\rm A}X_{\rm A} + \epsilon_{\rm B}(1-X_{\rm A})\right]C\right\}^2
\end{equation}
As the problem involves two unknowns, a minimum of two wavelengths is required for minimization. However, in order to obtain the highest accuracy, the fit was performed over a wide spectral bandwidth (420-1000~nm) using $\epsilon_{\rm A}$ and $\epsilon_{\rm B}$ as reference spectra. As discussed above, the calculation range was slightly narrower than the detection range of the spectrometer (345-1033 nm). The extreme wavelengths were excluded because the signal-to-noise ratio was below the detection limit of the instrument.

\begin{figure}[h!]
    \centering
    \includegraphics[scale=0.63]{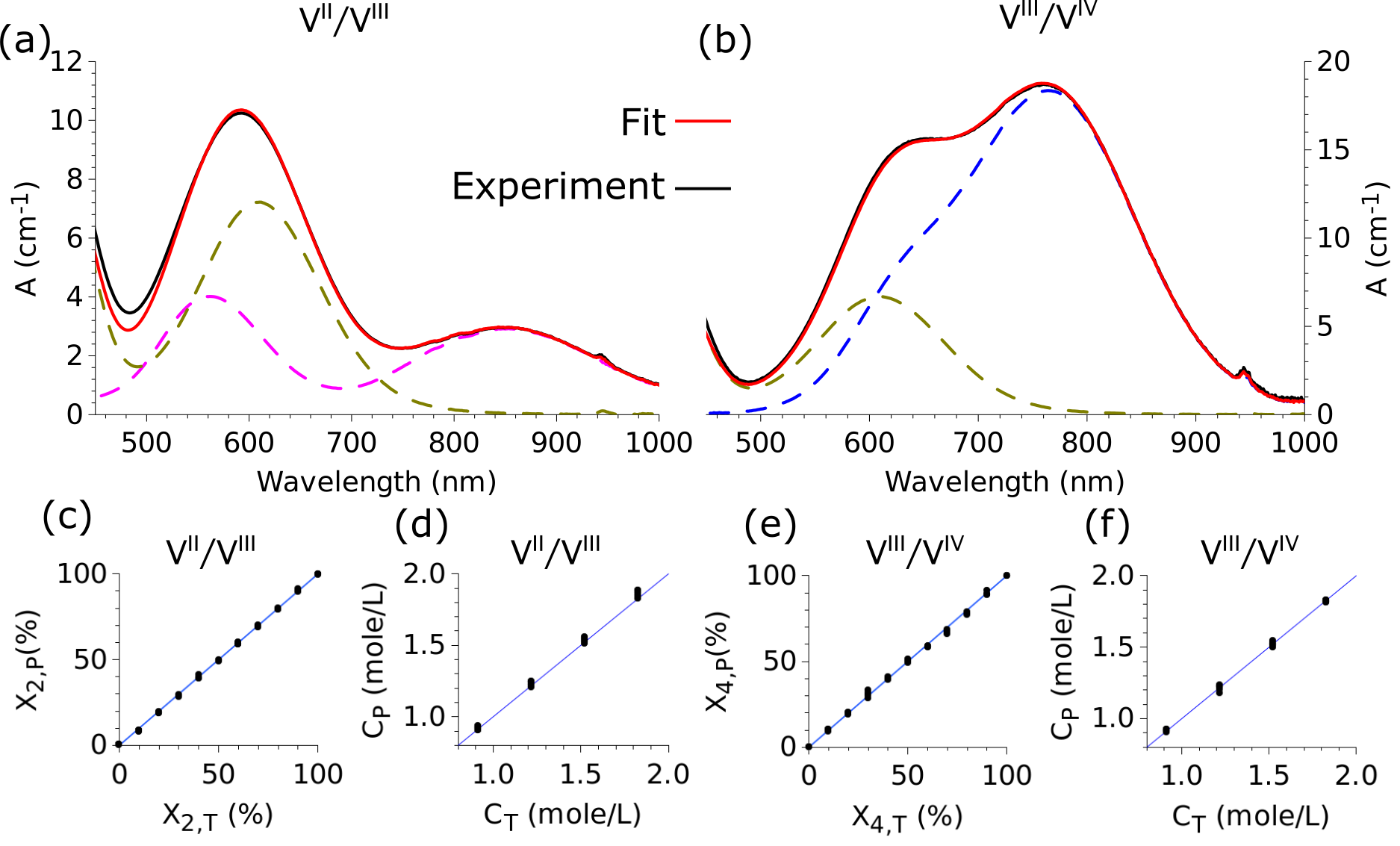}
    \caption{(a, b) Spectra of the $\VII$/$\VIII$ and $\VIII$/$\VIV$ mixtures for $X_{\rm A} = 50$\% and $C=1.83$~M. The dashed lines represent the spectral contributions of the pure compounds $\VII$: purple, $\VIII$: green, $\VIV$: blue. (c, d, e, f) Calibration curves representing the predicted values versus the true values for the mole fraction and the total concentration. The 100\% accuracy line ($y=x$) is plotted in blue.}
    \label{fig:lin_deconv}
\end{figure}

Loktionov et al.~\cite{loktionov_operando_2022} recently implemented this method for determining the SOC and the total concentration of vanadium ions, yielding solid results for vanadium concentrations of 0.5~M and 1~M. In this study, we extend the investigation to a broader concentration range and also incorporate the characterization of the $\VIII$/$\VIV$ mixture. Additionally, we report the measurement error when applying the method to calibrated samples. Figures~\ref{fig:lin_deconv}a and \ref{fig:lin_deconv}b illustrate the application of spectral deconvolution to the $\VII/\VIII$ and $\VIII/\VIV$ mixtures with two representative examples. The experimental absorbance spectrum for $C=1.83$~M and $X_{\rm A}=50$\%, with A denoting $\VII$ and $\VIV$, respectively, is shown in black. The dashed lines represent the spectral contributions of each pure compound: $\epsilon_{\rm A} X_{\rm A} C$ and $\epsilon_{\rm B} (1 - X_{\rm A})C$. According to Equation~\eqref{eq:deconv_linear}, the sum of the two dashed curves yields the red curve, which is the fit. Overall, the fit is accurate for all samples although a slight mismatch can sometimes be found at the lower wavelengths for the negative electrolyte, $\lambda < 600$~nm in Figure~\ref{fig:lin_deconv}a. This is consistent with the suggestion stated above that the spectral response of the mixture could lose linearity at lower wavelengths~\cite{loktionov_operando_2022}. 

For the established molar fraction and total concentration of each of the calibration samples, referred to as true values ($X_{\rm A,T}$, $C_{\rm T}$), spectral deconvolution was applied to obtain the so-called predicted values ($X_{\rm A,P}$, $C_{\rm P}$). Figures~\ref{fig:lin_deconv}c, \ref{fig:lin_deconv}d, \ref{fig:lin_deconv}e, and \ref{fig:lin_deconv}f show the calibration plots for $\VII$/$\VIII$ (c, d) and $\VIII$/$\VIV$ (e, f) for the total vanadium concentration $C$ and the mole fraction $X_A$. Essentially, these are scatter plots showing the predicted values versus the true values. Black dots represent the obtained results for all concentrations. In an ideal situation with perfect calibration and measurements with 100\% accuracy, the predicted values would align with the true values along the bisector of the first quadrant, represented by the blue line. In practical scenarios, deviations from the blue line indicate reduced calibration accuracy and increased measurement errors. To quantify accuracy, the RMSE was computed for all concentrations, as shown in Table \ref{tab:RMSE_deconv_lin}. The RMSE values were consistent over the entire concentration range, indicating uniform accuracy. Specifically, the average measurement errors for the mole fraction $X_{\rm A}$ and the concentration $C$ were found to be 0.79\% and 17 mM, respectively.

\begin{table}[h!]
    \centering
\begin{tabular}{ccccc}
  & \multicolumn{4}{c}{Electrolyte mixture} \\[1mm] \cline{2-5} \\[-3.5mm]
  & \multicolumn{2}{c}{$\VII$/$\VIII$} & \multicolumn{2}{c}{$\VIII$/$\VIV$} \\[1mm] \hline \\[-3.5mm]
$C$ (M) & $\RMSE_X$ (\%) & $\RMSE_C$ (M) & $\RMSE_X$ (\%) & $\RMSE_C$ (M) \\[1mm] \hline \\[-3.5mm]
1.83 & 1.11  & 0.038 & 0.65 &  0.009 \\
1.52 & 0.66  & 0.020 & 1.09 & 0.012 \\
1.22 & 0.75  & 0.019 & 0.52 &  0.015 \\
0.91 & 0.86  & 0.012 & 0.74 &  0.007 \\[1mm] \hline \\[-3.5mm]
 Average &\textbf{0.85}  & \textbf{0.022} & \textbf{0.74} & \textbf{0.012}
 \end{tabular}
\caption{Calibration results for the spectral deconvolution method showing the RMSE for the mole fraction in \% and the total concentration in~M for total vanadium concentrations between 0.91~M and 1.83~M.}
    \label{tab:RMSE_deconv_lin}
\end{table}

\subsection{Non-linear case: $\VIV/\VV$}

\begin{figure}[ht!]
    \centering
    \includegraphics[scale=0.63]{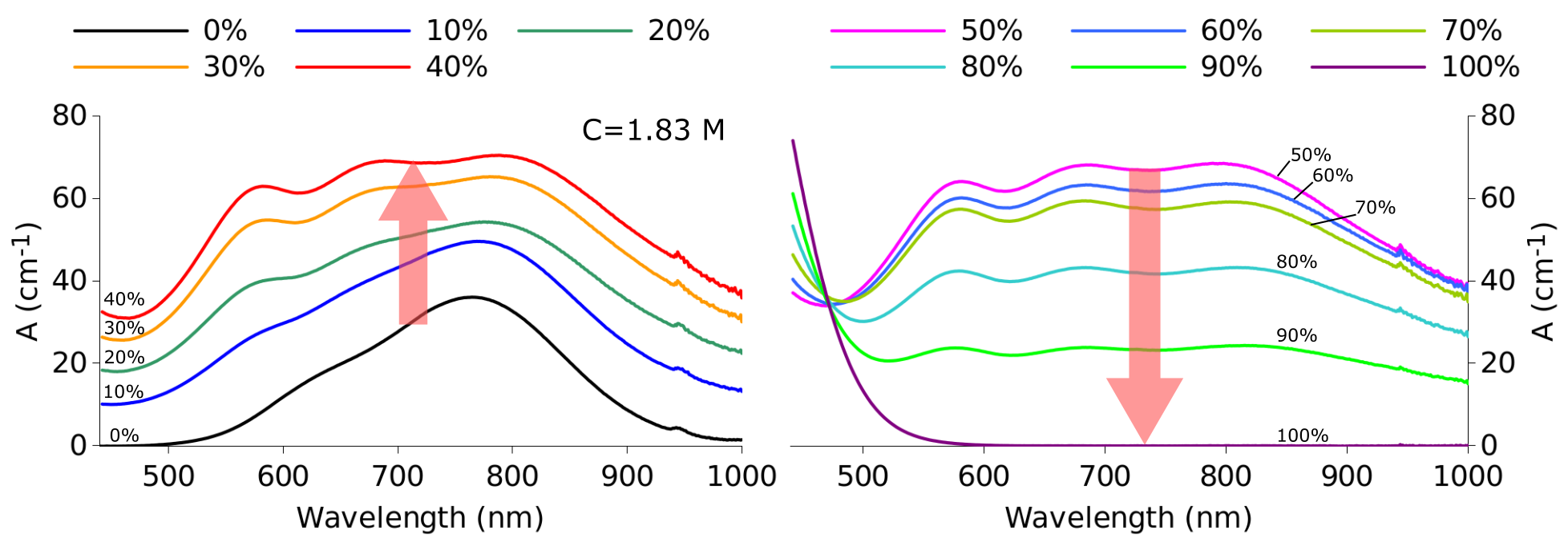}
    \caption{Absorbance spectra of the positive electrolyte $\VIV/\VV$ for different values of ${\rm SOC} = X_5 = 1 - X_4$, as indicated in the curve labels, and a total vanadium concentration of $C=1.83$~M. Note that above 600~nm the molar absorptivity of $\VV$ is virtually zero (see the bottom curve in the right subplot, $\rm SOC = 100\%$).}
    \label{fig:spectra_pos}
\end{figure}

Figure \ref{fig:spectra_pos} shows the absorbance spectra of the positive electrolyte $\VIV/\VV$ as the SOC increases from 0\% ($\VIV$) to 100\% ($\VV$) for a total concentration $C = 1.83$~M. As indicated by the red arrow, the overall absorbance (550-1000~nm) first increases up to ${\rm SOC} \approx 40\%$ (Figure \ref{fig:spectra_pos}a). Subsequently, the absorbance declines, first gradually within the 50-70\% SOC range and then sharply as the SOC approaches 80-100\% (Figure \ref{fig:spectra_pos}b). This observation reveals a radically different spectral response compared to that of $\VII/\VIII$ and $\VIII/\VIV$.
Another difference is that the catholyte $\VIV/\VV$ exhibits significantly higher absorbances, reaching values as high as 65~$\rm cm^{-1}$ over a wide range of wavelengths (560-850 nm) and SOCs (25-75\%).

According to Blanc et al.~\cite{blanc_spectrophotometric_1982}, the observed high-absorbance non-linear spectral behavior can be attributed to the formation of a 1:1 stoichiometric mixed-valence complex, $\rm V_2O_3^{3+}$. This complex is believed to form according to the equilibrium reaction
\begin{equation}
\label{eq:V2O33p_formation}
{\underbrace{\rm VO^{2+}\vphantom{_2}}_{\VIV} + \underbrace{\rm VO_2^{+}}_{\VV}} \; {\stackrel{K_c}{\rightleftharpoons}} {\rm \; V_2O_3^{3+}}    
\end{equation}
characterized by the equilibrium constant
\begin{equation} \label{eq:eqcondC4C4C45}
K_c = \frac{C_{45,\rm  }}{C_{4}C_{5}}    
\end{equation}
written here in terms of the equilibrium concentrations of $\VIV$, $\VV$, and that of the mixed-valence complex, denoted here as $C_{45}$. The overall stoichiometry of reaction \eqref{eq:V2O33p_formation} provides the following relations between the equilibrium concentrations
\begin{align}
C_4 & = C_4^*-C_{45} = X_4 C - C_{45} \label{eq:C4eq}\\
C_5 & = C_5^*-C_{45} = (1-X_4)C - C_{45} \label{eq:C5eq}
\end{align}
where $C_4^* = C_4 + C_{45}$ and $C_5^* = C_5 + C_{45}$ denote the nominal concentrations of $\VIV$ and $\VV$ in the absence of the mixed-valence complex, i.e., in the hypothetical case $K_c = 0$. In the above equations, $C_4^*$ and $C_5^*$ are conveniently rewritten as the product of the total vanadium concentration $C = C_4^*+C_5^* = C_4 + C_5 + 2C_{45}$ and the nominal mole fractions $X_4 = C_4^*/C$ and $X_5 = C_5^*/C$, which are in turn related to the SOC, given by $X_5 = 1 - X_4$. For further reference, the relations between $C_4$, $C_5$, $C_{45}$, $C$, $X_4$, $X_5$, and SOC are gathered in the last column of Table~\ref{tab:notation}.

Summarizing, the positive electrolyte contains 3 different compounds accounting for the total theoretical absorbance
\begin{equation}
    A_{\rm M} = A_{4} + A_{5} + A_{45}
\end{equation}
where $A_4$, $A_{5}$, and $A_{45}$ are the absorbances of $\VIV$, $\VV$, and $\rm V_2O_3^{3+}$, respectively. With use made of Equations~\eqref{eq:beer_lambert} and \eqref{eq:beer_lambert_power}, the absorbance of the mixture can be written as
\begin{equation}\label{eq:abs_pos}
        A_{\rm M} = \epsilon_4 C_{4} + \epsilon_5 C_{5}^k + \epsilon_{45} C_{45}
\end{equation}
Here we assume that the mixed valence complex $\rm V_2O_3^{3+}$ still follows the Beer-Lambert law~\eqref{eq:beer_lambert}, but generalize the analysis of Blanc et al.~\cite{blanc_spectrophotometric_1982} by considering 
the non-linear dependence between $A_5$ and the concentration of $\VV$ described by Equation\eqref{eq:beer_lambert_power}.

Combining Equations~\eqref{eq:C4eq}, \eqref{eq:C5eq}, and \eqref{eq:abs_pos}, we can write
\begin{equation}
\label{eq:abs_pos_2}
A_{\rm M} = \epsilon_4(X_4C-C_{45}) + \epsilon_5[(1-X_4)C-{C_{45}}]^k + \epsilon_{45}C_{45}
\end{equation}
Moreover, substituting Equations~\eqref{eq:C4eq} and \eqref{eq:C5eq} into the equilibrium condition~\eqref{eq:eqcondC4C4C45} yields a quadratic equation for $C_{45}$ that can be solved analytically to give~\cite{blanc_spectrophotometric_1982, quill_factors_2015}
\begin{equation}\label{eq:c45}
    C_{45} = \frac{1-\sqrt{1-4\chi^2X_4(1-X_4)C^2}}{2\chi}
\end{equation}
where
\begin{equation}\label{eq:chi}
   \chi = \frac{K_c}{K_cC + 1}
\end{equation}
According to the above analysis, during the charge and discharge of the $\VIV/\VV$ electrolyte the molar concentrations of the three compounds are expected to vary non-linearly with the state of charge, ${\rm SOC} = 1 - X_4$. This non-linear behavior poses a greater challenge for interpreting the spectra compared to the $\VII/\VIII$ and $\VIII/\VIV$ electrolytes.

It is useful to note that for wavelengths above 600~nm the molar absorptivity of $\VV$ is virtually zero, $\epsilon_5 \approx 0$, so that  Equation~\eqref{eq:abs_pos_2} reduces to
\begin{equation}
A_{\rm M}^{\lambda > 600{\rm nm}} =\epsilon_4(X_4 C-C_{45}) + \epsilon_{45}C_{45}
\end{equation}
which, upon substitution of $C_{45}$ from Equations~\eqref{eq:c45} and \eqref{eq:chi}, yields
\begin{equation}\label{eq:abs_pos_3}
A_{\rm M}^{\lambda > 600{\rm nm}} =\epsilon_4X_4C + (\epsilon_{45}-\epsilon_4)\left[\frac{1-\sqrt{1-4X_4(1-X_4)(K_cC)^2/(K_cC + 1)^2}}{2K_c/(K_cC + 1)}\right]
\end{equation}
In this equation, $K_c$ and $\epsilon_{45}$ are unknown constants, while $C$ and $X_4$ are known parameters for each  $\VIV/\VV$ calibration sample. Thus, since we have an experimental database of 44 reference spectra $A_{\rm exp}$ for different values of $C$ and $X_4$, we can perform a multivariate function fit to find the best pair ($\epsilon_{45}$, $K_c$) that, for each wavelength $\lambda > 600$~nm, minimizes the sum of the error squares
\begin{equation}
    \sum_{\underset{\lambda > 600{\rm nm}}{C,\hspace{1pt} X_4}} (A_{\rm exp} - A_{\rm M}^{\lambda > 600{\rm nm}})^2
\end{equation}
with the summation extending over all the 44 pairs ($C$, $X_4$) available in the $\VIV/\VV$ database.

\begin{figure}[t]
    \centering
    \includegraphics[scale=0.525]{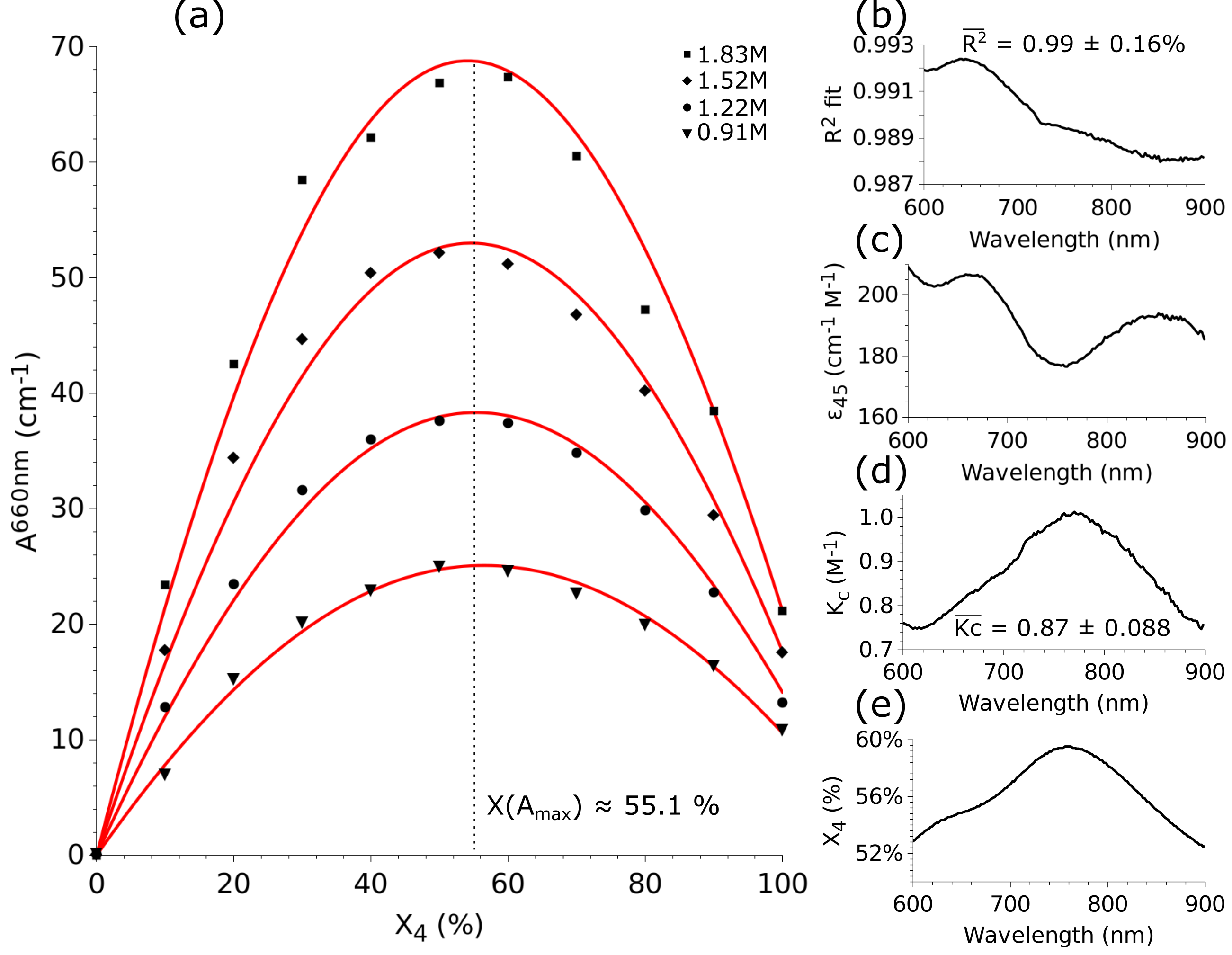}
    \caption{(a) Absorbance of the $\VIV/\VV$ mixture at 660~nm as a function of $X_4$ for the four total vanadium concentrations under study as measured experimentally (black dots) and fitted from Equation~\eqref{eq:abs_pos_3} (red curves). The right plots show the fitted (b) coefficient of determination $R^2$, (c) molar absorptivity of the complex, $\epsilon_{45}$, (d) equilibrium constant, $K_c$ along with the value of $X_4$ of maximum absorbance (e) as a function of the wavelength.}
    \label{fig:pos_fit}
\end{figure}

The black dots plotted in Figure~\ref{fig:pos_fit}a represent the experimentally measured absorbance at 660~nm, $A_{\rm exp}^{\rm 660nm}$, as a function of $X_4$. The red lines are the predicted absorbances resulting from the fitted values of $\epsilon_{45}$ and $K_c$ for that particular wavelength. Figure \ref{fig:pos_fit}b plots the coefficient of determination $R^2$ of the fit versus the wavelength. Details on the calculation of $R^2$ are given in the Supplementary Information (Section~S2). Overall, the model demonstrates excellent agreement with the experimental data, with a robust and stable value of $R^2$ averaging $\overline{R^2} \approx 0.99$ for $\lambda>600$~nm. The fitted values of $\epsilon_{45}$ and $K_c$ are represented in Figures~\ref{fig:pos_fit}c and \ref{fig:pos_fit}d as a function of the wavelength. It is important to clarify that the observed variation of approximately $\pm 10\%$ in $K_c$ is either a numerical artifact that results from the multivariate regression process or could be attributed to model limitations and experimental errors. The reaction constant $K_c$ is a chemical property of the electrolyte, and therefore it should remain constant regardless of the wavelength sampled.

The average value of $K_c$ indicated in the plot ($\overline{K}_c = 0.87 \pm 0.088~\rm M^{-1}$) falls within the higher end of the range 0.2-0.8 $\rm M^{-1}$ reported in the literature~\cite{blanc_spectrophotometric_1982, buckley_towards_2014, quill_factors_2015, loktionov_operando_2022}. Although the discrepancy is small, it could be traced back to the lower total vanadium concentrations used in the cited experiments. Additionally, by expressing the absorbance of the isolated complex in the form
\begin{equation}
A_{45} = \epsilon_{45}C_{45} = \epsilon_{45} K_c C_4C_5
\end{equation}
it becomes clear that there is an inverse relationship between $K_c$ and the molar absorptivity $\epsilon_{45}$, so that the fitting could perform equally well for different pairs ($\epsilon_{45}$, $K_c$) just by increasing $K_c$ and reducing $\epsilon_{45}$. Please note that the values of $K_c$ reported in the literature have always been obtained indirectly, as we do in this study, without direct measurements of the complex concentration.

The last subplot, Figure~\ref{fig:pos_fit}e, shows $X_4(A_{\rm max})$ as a function of the wavelength. This is essentially the value of $X_4$ that gives maximum absorbance for a given wavelength, e.g., $X_4 = 55.1\%$ for $\lambda = 660$~nm, as shown in Figure~\ref{fig:pos_fit}a. This turning point is of practical importance as it serves as a reference marker for estimating the total concentration between charging cycles. 

The results presented in this section are used below to define two calibration methods allowing to estimate the ${\rm SOC} = 1 - X_4$ and the total vanadium concentration $C$ of the catholyte $\VIV/\VV$.

\subsubsection{Fast empirical method}

The nonlinear character of Equation~\eqref{eq:abs_pos_3} makes it impossible to solve it analytically for either $X$ or $C$. An alternative approach involves fitting the absorbance of the mixture to a quadratic function of $X_4$ and $C$ of the form
\begin{equation}\label{eq:abs_pos_empir}
    A_{\rm M} = a_0X_4C + a_1X_4C^2 + a_2X_4^2C + a_3X_4^2C^2
\end{equation}
Table~\ref{tab:param_pos_direct} lists the values of the fitting parameters $a_i$ for two selected wavelengths in the range $\lambda > 600$~nm. 
The reported values were obtained through multivariate polynomial regression of the experimental absorbance data from the 44 calibration samples of the $\VIV/\VV$ mixture at $660$~nm and $760$~nm. 
As demonstrated in the Supplementary Information (Figure~S2), the quadratic fit \eqref{eq:abs_pos_empir} exhibits comparable performance to the theoretical equation \eqref{eq:abs_pos_3} with an average $R^2$ 
of 0.99.

\begin{table}[hb]
    \centering
\begin{tabular}{ccc}
& \multicolumn{2}{c}{Wavelength $\lambda$} \\[1mm]
\cline{2-3} \\[-3.5mm]
  Fitting parameter  & 660 nm & 760 nm \\[1mm]
\hline \\[-3.5mm]
$a_0$ ($\rm M^{-1}$) & $\phantom{-}62.12$ & $\phantom{-}71.29$ \\
$a_1$ ($\rm M^{-2}$)& $\phantom{-}41.83$ & $\phantom{-}33.62$ \\
$a_2$ ($\rm M^{-1}$)& $-50.65$ & $-51.71$ \\
$a_3$ ($\rm M^{-2}$)& $-42.63$ & $-34.56$ \\
\end{tabular}
\caption{Values of the fitting parameters $a_i$ appearing in Equation~\eqref{eq:abs_pos_empir} obtained from the absorbances of the 44 calibration samples of the $\VIV/\VV$ mixture at two selected wavelengths in the range $\lambda > 600$~nm.}
    \label{tab:param_pos_direct}
\end{table}

\begin{figure}[t!]
    \centering
    \includegraphics[scale=0.63]{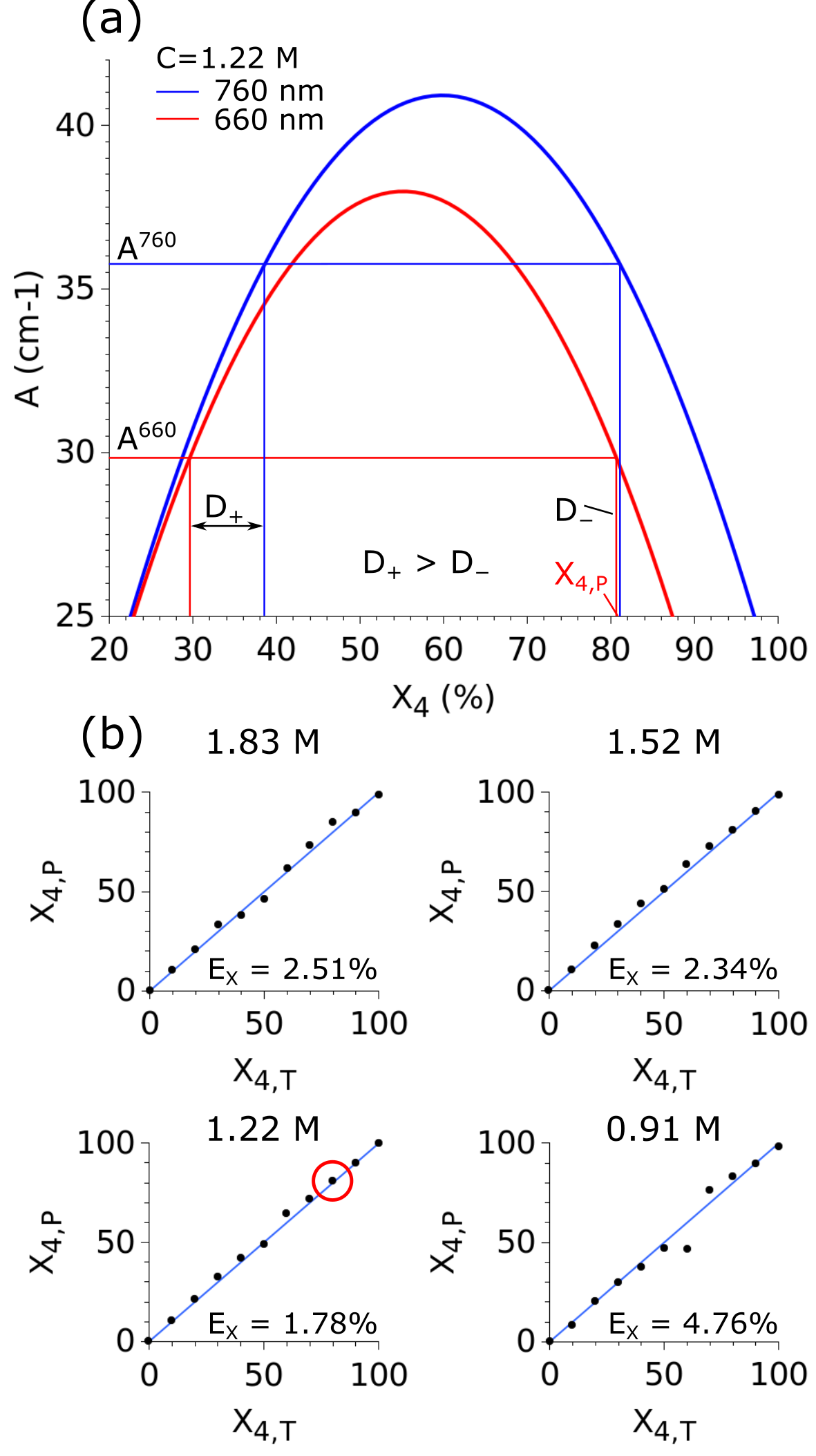}
    \caption{(a) Absorbance versus $X_4$ curves computed for $C=1.21$~M using the quadratic fit~\eqref{eq:abs_pos_empir} at 760~nm (blue) and 660~nm (red). The horizontal and vertical lines act as sample guidelines for the calculation of $D_-$, $D_+$, and $X_{\rm 4,P}$, all labeled in the graph. (b) Calibration results (black dots) obtained using the fast empirical method for the four concentrations under study. The 100\% accuracy line ($y=x$) is plotted in blue. The dot highlighted in red for $C = 1.22$~M corresponds to the sample calculation shown in (a)}
    \label{fig:pos_direct}
\end{figure}

If the total vanadium concentration $C$ is known, then $X_4$ can be obtained  analytically by solving the quadratic Equation~\eqref{eq:abs_pos_empir} to give
\begin{equation}\label{eq:root_X}
    X_4 =  \frac{-(a_0C + a_1C^2) \pm \sqrt{(a_0C + a_1C^2)^2+4A_{\rm M}(a_2C+a_3C^2)}}{2(a_2C+a_3C^2)}
\end{equation}
which yields two possible roots ($X_{4-}$, $X_{4+}$). However, one of these roots is spurious. To elucidate which is the correct value of $X_4$ it is necessary to use Equation~\eqref{eq:root_X} with two different wavelengths ($\lambda_1$, $\lambda_2$). By comparing the solutions obtained with these two wavelengths, the false value of $X_4$ can be disregarded~\cite{petchsingh_spectroscopic_2016}. Figure \ref{fig:pos_direct} illustrates the process, based on evaluating the squared differences between the roots $X_{4-}$ and $X_{4+}$ provided by the two wavelengths
\begin{align}
    D_- & = \left(X^{\lambda_1}_{4-} - X^{\lambda_2}_{4-}\right)^2 \\
    D_+ & = \left(X^{\lambda_1}_{4+} - X^{\lambda_2}_{4+}\right)^2
\end{align}
and selecting the root that yields the smallest difference. The value of $X_4$ is then computed as the arithmetic mean of the selected root obtained from both wavelengths.

As an illustrative example, Figure \ref{fig:pos_direct}a represents the absorbance versus $X_4$ curves computed using Equation~\eqref{eq:abs_pos_empir} with $C~=~1.22$~M and the fitting parameters reported in Table~\ref{tab:param_pos_direct} for 660~nm and 760~nm. An illustrative example demonstrating the application of the process for determining the composition $X_{4,P}$ of a particular mixture is also provided. In this case, we used the calibration sample corresponding to $C = 1.21$~M and $X_{4,\rm T} = 80\%$, plotting the experimentally measured absorbances $A_{\rm M}^{\rm 660nm}$ and $A_{\rm M}^{\rm 760nm}$ as horizontal lines. The intersection points of these lines with the parabolic curves represent the potential values of $X_{4,\rm P}$. Upon analysis, it is observed that $D_-<D_+$, indicating that the correct root corresponds to the one associated with the $-$ sign. Thus, the value of $X_{4,\rm P}$ can be obtained by averaging, in this case $X_{4,\rm P} = (X_{4-}^{\rm 660nm} + X_{4-}^{\rm 760nm})/2 = {\rm 80.9\%}$.

Figure \ref{fig:pos_direct}b shows the calibration curves representing the predicted ($X_{\rm 4,P}$) versus the true ($X_{\rm 4,T}$) values of $X_4$ for each of the four concentrations within the 44 calibration samples of the $\VIV/\VV$ mixture. 
The error $\RMSE_X$ is also indicated in the plots. The red circle located in the 1.22~M curve corresponds to the calibration data used as the example in Figure~\ref{fig:pos_direct}a.
It is worth noting that the method presented here is weaker than the one proposed in Section~\ref{sec:empirical234} for the $\VII/\VIII$ and $\VIII/\VIV$ mixtures, since in this case $X_4$ and $C$ cannot be obtained at once using only two absorbance values. Indeed, the total concentration must be either known \textit{a-priori} or assumed, which may give rise to systematic errors over the long-term cycling of the battery due to the accumulated effect of ion crossover.

One possible way to determine the total vanadium concentration is to monitor the absorbance of the $\VIV/\VV$ mixture during cycling in real time. This enables the estimation of $C$ from the peak absorbance, $A_{\rm max}$, observed during the charge or discharge process. As previously discussed, the mole fraction at the turning point, $X_4(A_{\rm max})$, is well known for all wavelengths (see Figure \ref{fig:pos_fit}e). Thus, by using the pair $(A_{\rm M}, X_4) = (A_{\rm max}, X_4(A_{\rm max}))$ one can estimate the total concentration by solving the quadratic Equation~\eqref{eq:abs_pos_empir} to give
\begin{equation}\label{eq:root_C}
    C =  \frac{-(a_0X_4 + a_2X_4^2) + \sqrt{(a_0X_4 + a_2X_4^2)^2+4A_{\rm M}(a_1X_4+a_3X_4^2))}}{2(a_1X_4+a_3X_4^2)}
\end{equation}

\subsubsection{Spectral deconvolution}

\begin{figure}[t!]
    \centering
    \includegraphics[scale=0.63]{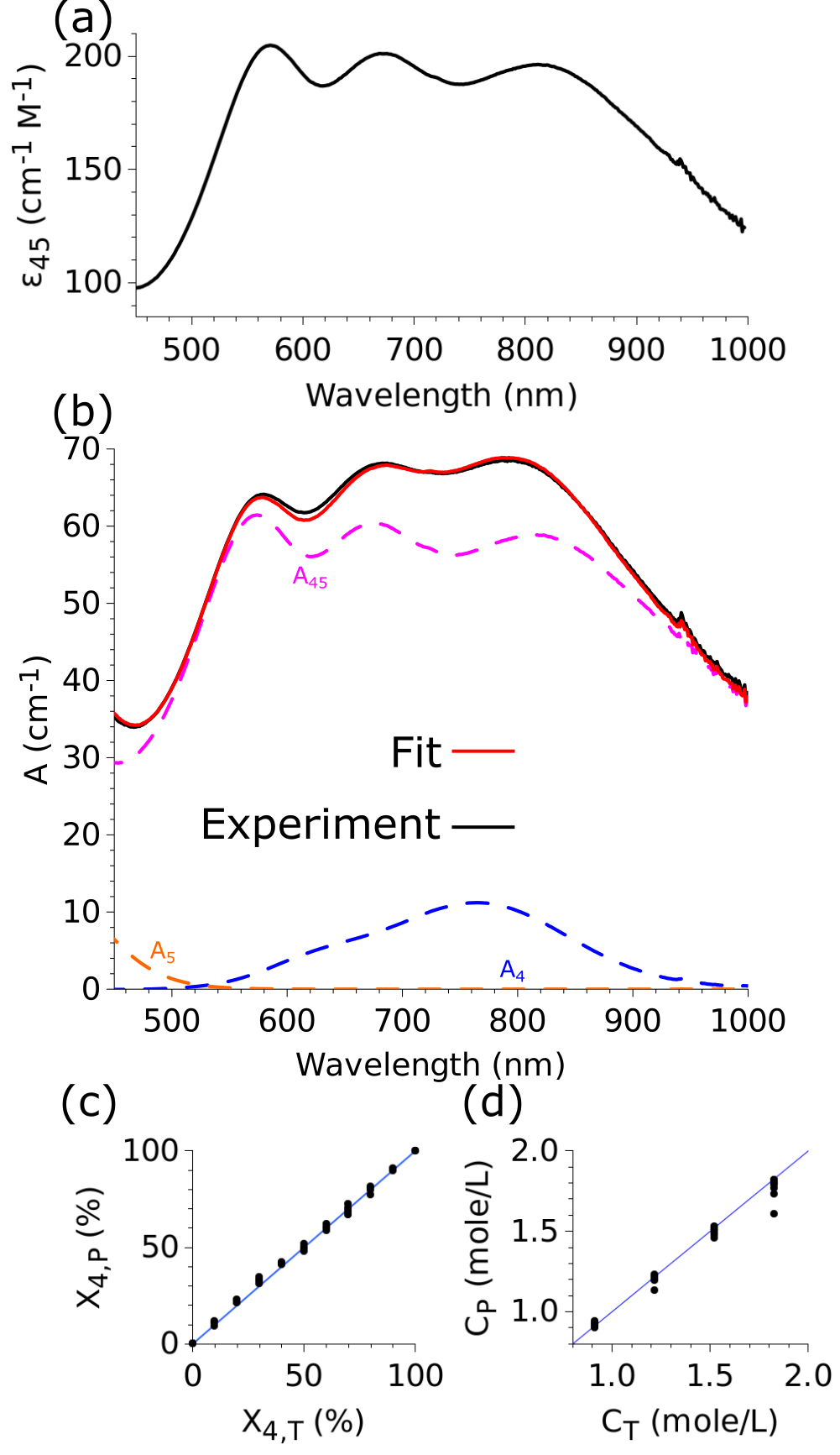}
    \caption{(a) Absorptivity spectrum of the complex $\rm V_2O_3^{3+}$ provided by Equation~\eqref{eq:e45}; (b) deconvolution results for $X_4 = 50\%$ at 1.83M, with the spectral contributions of the three compounds, $A_4$, $A_5$, and $A_{45}$, shown as dashed lines; and calibration curves obtained using the spectral deconvolution method for (c) $X_4$ and (d) $C$, with the 100\% accuracy line ($y=x$) plotted in blue.}
    \label{fig:pos_deconv}
\end{figure}

This method is similar to the one presented in Section~\ref{sec:deconv_lin}, but the equations here are slightly different due to the presence of the complex $\rm V_2O_3^{3+}$. The first requirement is to determine the molar absorptivity spectrum of the complex $\epsilon_{45}$. Since this species does not appear in isolation from $\VIV$ and $\VV$, the experimental determination of $\epsilon_{45}$ must be done indirectly. To achieve this, we can rewrite Equation~\eqref{eq:abs_pos_2} as
\begin{equation}\label{eq:e45}
    \epsilon_{45} = \frac{A_{\rm exp} - \epsilon_4(X_4C-{C_{45}}) - \epsilon_5\left[(1-X_4) C-C_{45}\right]^k} { C_{45}}
\end{equation}
where $C_{45}$ is evaluated from Equation~\eqref{eq:c45} using the approximate value of $\chi \approx \overline{K}_c/(\overline{K}_c + 1)$ obtained from Equation~\eqref{eq:chi} with the average value of the equilibrium constant reported above, $\overline{K}_c=0.87 \rm~M^{-1}$. The equation is applied for all the calibration samples (36 in total, disregarding the 8 samples with either $X_4$ or $X_5 = 0$) using $A_{\rm exp}$ as the measured absorbance. The reference $\epsilon_{45}$ curve, plotted in Figure \ref{fig:pos_deconv}a, is computed by averaging over all the calibration samples (see Figure~S3). Further details regarding the averaging of $\epsilon_{45}$ and the limits of the model are given in the Supplementary Information (Section~S3). The high molar absorptivity of $\rm V_2O_3^{3+}$ is noteworthy, particularly when compared to those of $\VIV$ and $\VV$; e.g., while $\epsilon_{45}$ reaches values as high as $200~\rm cm^{-1}M^{-1}$, $\epsilon_4$ peaks at about $20~\rm cm^{-1} M^{-1}$ (see Figure~\ref{fig:spectra_soc})

Once the absorptivity spectrum $\epsilon_{45}$ is known, we can apply the spectral deconvolution method to any arbitrary mixture of positive electrolyte $\VIV/\VV$. The procedure involves determining the pair ($X_4$, $C$) that minimizes the sum of the error squares between the measured mixture absorbance $A_{\rm exp}$ and the absorbance predicted by Equation~\eqref{eq:abs_pos_2}, that is,
\begin{equation}
\label{eq:deconv_pos}
    \sum_{\lambda} \left\{A_{\rm exp} - \epsilon_4(X_4C - C_{45}) - \epsilon_5\left[(1-X_4)C- C_{45}\right]^k - \epsilon_{45} C_{45}\right\}^2
\end{equation}
where the summation ranges from 440~nm to 1000~nm, the widest wavelength range that the spectrometry system offers with a sufficient signal-to-noise ratio for the positive electrolyte.

The method proposed here includes several enhancements with respect to that proposed by Loktionov et al.~\cite{loktionov_operando_2022}. A key innovation in our case is the incorporation of the power-law absorbance versus concentration response for $\VV$ stated in Equation~\eqref{eq:beer_lambert_power}, which translates into the nonlinear term on the right hand side of Equation~\eqref{eq:abs_pos_2}. This generalization significantly improves the calibration results. Indeed, the use of $k=1$ in Equation~\eqref{eq:beer_lambert_power} leads to poor fitting results for the $\VV$ spectrum at wavelengths lower than 600~nm.
A parametric study exploring various values of $k$ is provided in the Supplementary Information (Section~S4), demonstrating that the optimal power-law exponent is $k = 2.09$. This is close to the value $1.88$ reported in Figure \ref{fig:ref_spectra}, fitted using only one wavelength.
Additionally, our method introduces other calibration improvements. By considering the entire range of wavelengths instead of dividing it into two ranges, we eliminate one fitting step. Moreover, the use of the analytical expressions \eqref{eq:abs_pos_2} and \eqref{eq:c45} directly provides the desired values of $C$ and $X_4$ for a broader range (0.91-1.83~M) of total vanadium concentrations.

Figure \ref{fig:pos_deconv}b provides an illustrative example of the spectral deconvolution process showing the individual contribution of all compounds. The agreement is excellent, with the red curve (fitted) accurately overlapping the black curve (measured). Figures \ref{fig:pos_deconv}c and \ref{fig:pos_deconv}d show the calibration curves for $X_4$ and $C$. Finally, Table \ref{tab:RMSE_deconv_pos} indicates the RMSE errors resulting from the application of the spectral deconvolution method to both variables. 

\begin{table}[htbp]
    \centering
\begin{tabular}{ccc}
$C$ (M) & $\RMSE_X$ (\%) & $\RMSE_C$ (M) \\[1mm]
\hline \\[-3.5mm]
1.83 & 2.26  & 0.078  \\
1.52 & 1.70  & 0.028  \\
1.22 & 1.07  & 0.029 \\
0.91 & 1.34  & 0.013 \\[1mm]
\hline \\[-3.5mm]
 Average &\textbf{1.59}  & \textbf{0.037}
 \end{tabular}
\caption{RMSE error of the spectral deconvolution method of the positive electrolyte $\VIV/\VV$ corresponding to the $\VIV$ mole fraction, $X_4$, and the total vanadium concentration, $C$.}
    \label{tab:RMSE_deconv_pos}
\end{table}

\section{Discussion}
\label{sec:discussion}

\begin{figure}[t!]
    \centering
    \includegraphics[scale=0.63]{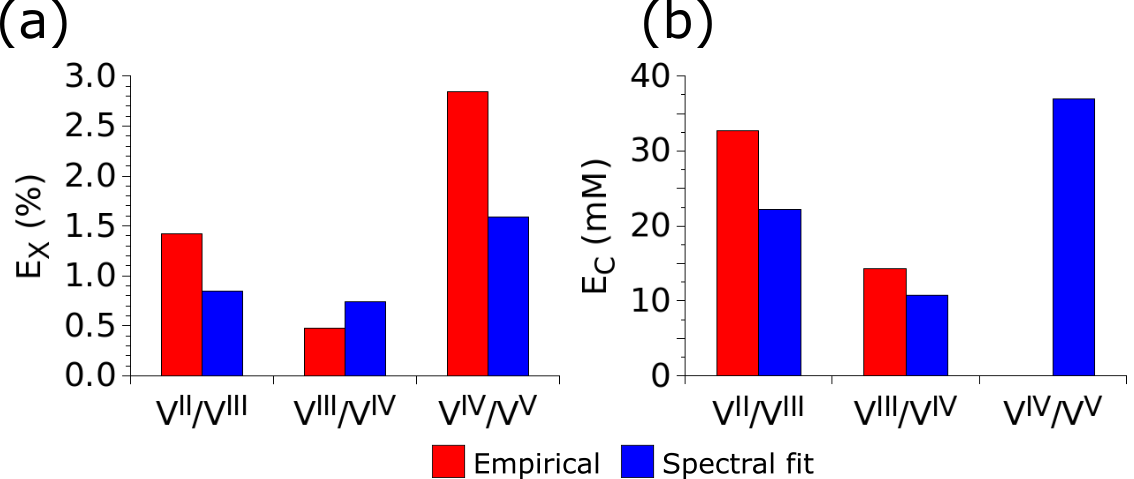}
    \caption{Histogram plots representing the average RMSE errors (a) $\RMSE_X$ and (b) $\RMSE_C$ for the three vanadium mixtures under study corresponding to the empirical (red) and spectral deconvolution (blue) methods.}
    \label{fig:summary_sigma}
\end{figure}

Figure \ref{fig:summary_sigma} shows histogram plots of the average RMS errors $\RMSE_X$ and $\RMSE_C$ for the three vanadium mixtures under study, comparing the empirical and spectral deconvolution methods.
Among the various vanadium mixtures, the calibrations perform slightly worse for the positive electrolyte $\VIV$/$\VV$. This is expected because the spectral response is significantly altered by the presence of the high absorbance complex $\rm V_2O_3^{3+}$, as discussed in the previous section. On average, the spectral deconvolution method performs better in both measuring the total concentration $C$ and the representative vanadium mole fractions $X_i$, $i = \{2, 4\}$. Additionally, no red bar is plotted for the $\RMSE_C$ error of the the positive electrolyte since the empirical calibration measures only $X_4$. 
The two calibration approaches differ radically in the quantity of information given to the algorithm. One the one hand, the empirical method is designed to use only a few hand-picked values while the spectral deconvolution uses the entirety of the spectrum. The deconvolution method essentially takes advantage of the whole spectrum to extract the maximum amount of information contained in it. We also expect the deconvolution approach to be more resilient to spectral distortions due to experimental errors. Such distortions can occur due to poor referencing, contaminated flow cuvettes (e.g., particles, bubbles), light scattering, change of light intensity over time, etc. These induces the spectra to be distorted or offsetted. We observed experimentally that, in such poor conditions the deconvolution was somehow able to absorb the distortions and delivered very acceptable values while the empirical methods yielded incorrect values. To delve further and assess this resilience, it would be useful to purposefully induce these distortions in a controlled manner to test the robustness of the calibration methods, but this is outside the scope of this paper. Moreover, other calibration methods could be envisioned such as correlation analysis, principal component regression, partial least square as well as advanced machine learning, such methods being already applied in other chemometrics studies~\cite{liu_state_2012, guo_chemometric_2021, meza_ramirez_applications_2021}.

\section{Conclusions}
\label{sec:conclusions}

This study presents an exhaustive overview of UV-Vis absorbance spectroscopy calibration applied to vanadium electrolytes to estimate the total vanadium concentration $C$ and the nominal mole fraction of the mixture compounds $X_i$. We considered the three known mixtures of vanadium electrolytes: $\rm V^{2+}/V^{3+}$ ($\VII/\VIII$), $\rm V^{3+}/VO_2^{+}$ ($\VIII/\VIV$), and $\rm VO_2^{+}/VO^{2+}$ ($\VIV/\VV$).
In total, we provide 6 calibration methods, 3 empirical while the other 3 require computer-based fits based on spectral deconvolution. On average, we obtained errors for the mole fraction $\RMSE_X \approx 1.0$-$1.5 \%$ and for the total concentration $\RMSE_C \approx 25$-$35$~mM.
Using a spectrum as the input, almost all calibrations provide both $C$ and $X_i$ as an output for a wide range of concentration (0.91-1.83M).

The empirical calibrations are aimed to be used in industrial applications, where a low-cost optical sensor picking up only two absorbance values could measure the SOC of the battery in real-time. This would supposedly come with a slightly worse accuracy but at a lower cost. The spectral deconvolution method is considered more powerful but computationally intensive, so it is better suited for laboratory applications where spectrometers and computers are common, but comes at a higher cost.

The data provided with this paper is publicly shared in open access repositories (\href{https://github.com/AngeAM/SOC_Vanadium_Spectra_2023.git}{GitHub}) along with the Python code used to perform the calibrations. As such, anyone collecting spectral data of their vanadium electrolyte could measure the state of charge and the concentration of their electrolytes using our code and data. The main idea of this study is to provide an exhaustive, clear and useful guide to both industry and academia practitioners on how to accurately measure the total and relative concentrations of the different species present in vanadium electrolytes using UV-Vis absorbance spectroscopy.
This study brings valuable insights into the constantly evolving body of knowledge on vanadium electrolytes, which will hopefully lead to the design and development of more efficient and competitive vanadium redox flow batteries in the near future.

\section*{Data availability}
The calibration data and software tools developed in this work are publicly shared in the open repository \href{https://github.com/AngeAM/SOC_Vanadium_Spectra_2023.git}{https://github.com/AngeAM/SOC\_Vanadium\_Spectra\_2023.git} under the open access MIT license system. Detailed information on the metadata and the calibration scripts are given in the readme file located at the root of the repository. 

\section*{Acknowledgments}

We would like to acknowledge the critical hardware support provided by Micro Electrochemical Technologies S.L. (B5tec). We would also like to thank Alberto Bernaldo for his support in setting up the experimental platform, Sonia Sevilla for the help in preparing the samples, and Dr.~Vanesa Muñoz-Perales for the design of the electrochemical cell. Finally, we acknowledge the support of Jesús Palma from IMDEA energía for the ICP analysis and the electrolyte supply.

\section*{Author contributions}
Ange A. Maurice: Conceptualization, Methodology, Software, Validation, Formal analysis, Investigation, Resources, Data Curation, Writing - Original Draft, Visualization, Project administration, Funding acquisition. Alberto E. Quintero: Conceptualization, Methodology, Investigation, Resources, Writing - Original Draft, Supervision. Marcos Vera: Conceptualization, Formal analysis, Investigation, Resources, Writing - Review \& Editing, Supervision, Project administration, Funding acquisition.

\section*{Funding}
This work has been partially funded by FEDER/MICINN-
Agencia Estatal de Investigación under projects TED2021-129378B-C21 and PID2019-106740RB-I00/AEI/10.13039/501100011033. A.A.~Maurice acknowledges the support of an MCSF-Cofund “Energy for Future” (E4F) postdoctoral research fellowship by the Spanish Iberdrola Foundation (GA-101034297).

% Bibliography
\bibliographystyle{elsarticle-num} 
\bibliography{references.bib}

% Auxiliar. Cuando termine lo quito
%\input{Supplementary_material.tex}

\end{document}

\begin{frontmatter}

\title{\textbf{Supplementary Information to} \\[8mm] 
A comprehensive guide for measuring total vanadium concentration and state of charge of vanadium electrolytes using UV-Visible spectroscopy}
\date{}

\author[aff1]{Ange A. Maurice}
\author[aff1,aff2]{Alberto E. Quintero}
\author[aff1]{Marcos Vera}

\affiliation[aff1]{organization={Departamento de Ingeniería Térmica y de Fluidos, Universidad Carlos III de Madrid},%Department and Organization
            addressline={Avd. de la Universidad 30}, 
            city={Leganés},
            postcode={28911}, 
            state={Madrid},
            country={Spain}}

\affiliation[aff2]{organization={R\&D Department, Micro Electrochemical Technologies},%Department and Organization
            city={Leganés},
            postcode={28918}, 
            state={Madrid},
            country={Spain}}

%\begin{abstract}

%\end{abstract}

\end{frontmatter}

\renewcommand{\thepage}{S\arabic{page}}
\renewcommand{\thesection}{S\arabic{section}}
\renewcommand{\thetable}{S\arabic{table}}
\renewcommand{\thefigure}{S\arabic{figure}}

\section{Preparation of $\VIII$ and $\VIV$ via spectrophotometric titration}
This part describes the preparation of pure $\VIII$ and $\VIV$ through spectrophotometric titration outlined in Section 2.1.
Spectrophotometric titration is a quantitative analytical technique that combines the principles of spectrophotometry and titration. This technique involves measuring the absorbance of a sample at specific wavelengths to determine the concentration of a particular substance. In the case of vanadium electrolytes, the titration process involves the gradual oxidation or reduction of the sample, which alters its light-absorbing properties. 

Figure S1 illustrates the application of spectrophotometric titration in monitoring the charging process of the $\VIII/\VIV$ electrolyte. The absorbance $A^\lambda$ is plotted as a function of time at specific wavelengths both for the positive side (oxidation, $\lambda=470$ nm) in Figure S1a and the negative side (reduction, $\lambda=850$ nm) in Figure S1b. The inflection points observed in the curves indicate the precise moment when the electrolyte consists solely of one oxidation state, signifying the transition state, or endpoint, from one electrolyte type to another. These distinct minima serve as reference points to prepare the reference electrolytes, enabling accurate analysis and comparison. 

\begin{figure}[h!]
    \centering
    \includegraphics[scale=0.5]{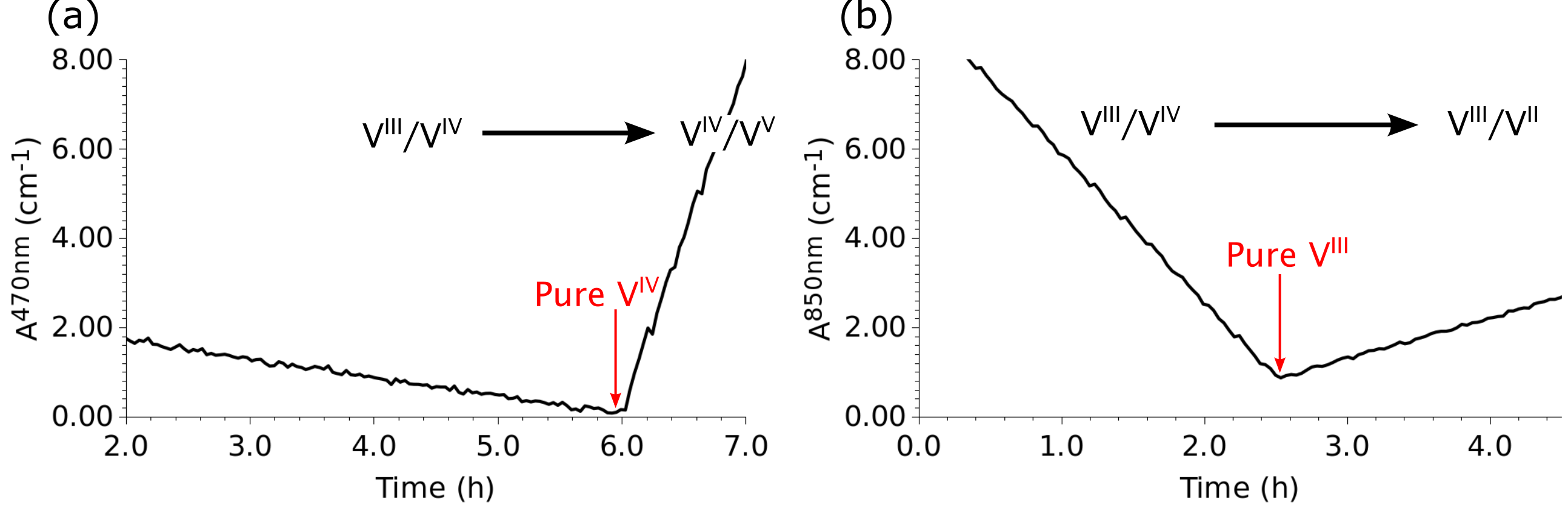}
    \caption{Plots of the absorbance $A^\lambda$ as a function of time at specific wavelengths during the charging process of the $\VIII/\VIV$ electrolyte for (a) the positive side and (b) the negative side. A distinct inflection point can be observed in each subplot indicating the transition to a different electrolyte mixture.}   
\end{figure}

\section{Empirical fitting of the positive electrolyte and calculation of $R^2$}
This part details the fitting of the positive electrolyte absorbance using the empirical method outlined in Section 3.3.1 as well as the calculation of $R^2$ (also used in 3.3, Figure 6b).
Figure S2a shows the absorbance of $\VIV/\VV$ at 660~nm as a function of $X_4$ for the four concentrations. The black dots are the experimental points ($A_{\rm exp}$) and the red line is the fitted curve ($A_{\rm p}$) obtained from Equation (22). As pointed out in the main text, the empirical fit performs as well as the analytical fit shown in Figure 6. This is further confirmed by the high value of $R^2$ shown in Figure S2b.
The coefficient of determination is calculated at each wavelength following its classical definition
$$R^2 = 1 - {SS_{\rm res}\over SS_{\rm tot}} $$
in terms of the sum of squares of residuals
$$SS_\text{res}=\sum_{C,X_4} (A_{\rm exp} - A_{\rm p})^2$$
where $A_{\rm p}$ denote the predicted absorbance values (plotted as red lines in Figure 6a and Figure S2a), and the total sum of squares
$$SS_\text{tot}=\sum_{C, X_4} (A_{\rm exp} - \overline{A_{\rm exp}})^2$$
defined as the sum over all squared differences between the observations and their overall mean.
\begin{figure}[h!]
    \centering
    \includegraphics[scale=0.5]{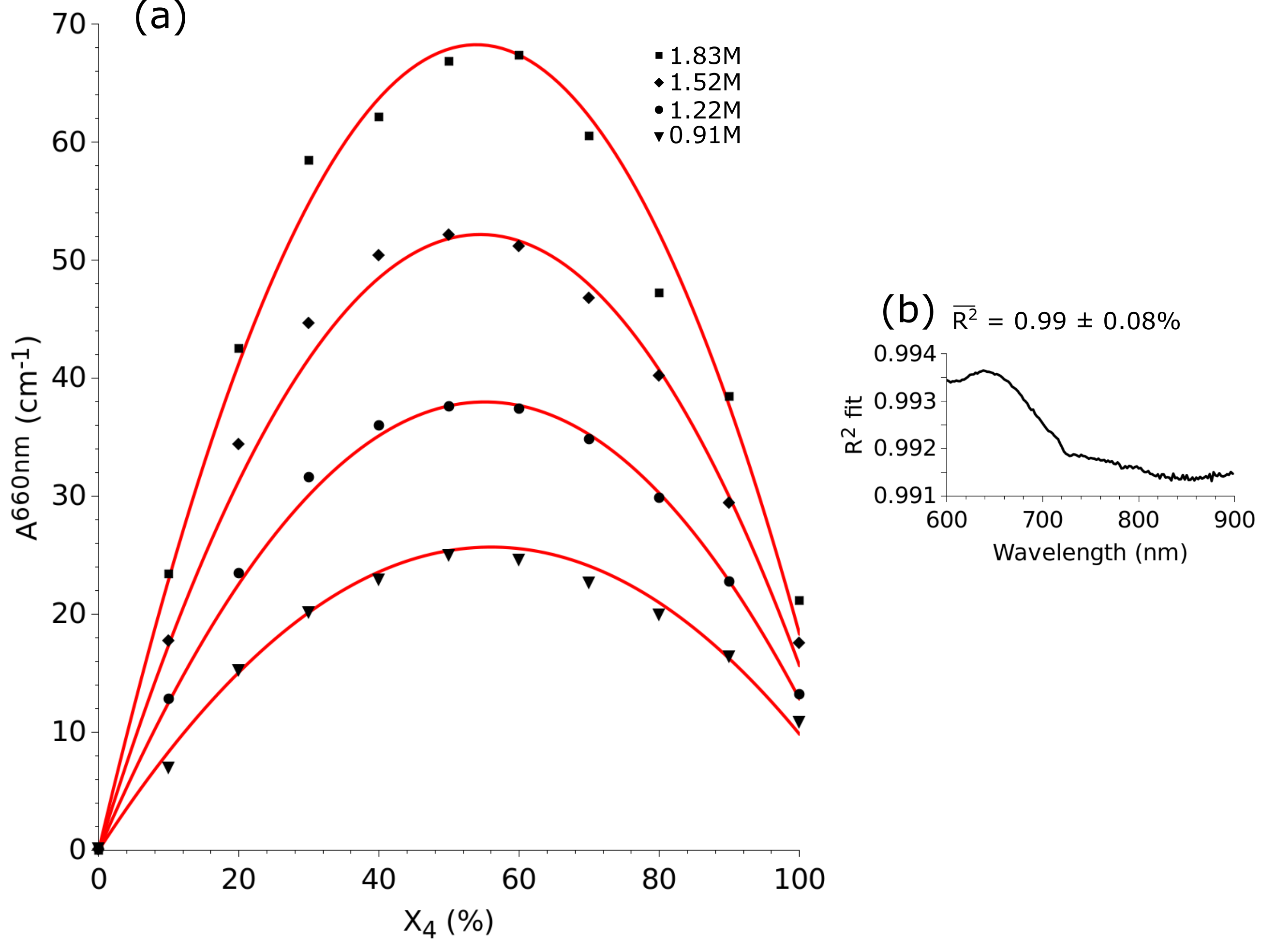}
     \caption{(a) Absorbance of the $\VIV/\VV$ mixture at 660~nm as a function of $X_4$ for the four total vanadium concentrations under study as measured experimentally (black dots) and fitted from the empirical Equation~22 (red curves). The right plot show the fitted (b) coefficient of determination $R^2$ as a function of the wavelength.}
\end{figure}

\section{Averaging of $\epsilon_{45}$}

This section describes the procedure for determining the molar absorptivity of the complex, $\epsilon_{45}$, used in the spectral deconvolution of the positive electrolyte as described in Section 3.3.2 of the main text.
Figure S3 shows in gray the calculated absorptivities of the complex $\epsilon_{45}$ obtained from Equation~(27) for all 36 calibration samples of the positive electrolytes. In order to exclude samples where the complex is absent, those with $X_4 = 0$\% or 100\% were omitted.
According to the model assumptions, only one unique $\epsilon_{45}$ curve should be associated with the complex, regardless of the values of $X$ or $C$. The average of the gray curves is arbitrarily chosen as a reference (depicted as the red curve) for the calibration process, yielding satisfactory deconvolution results as highlighted in the main text (with a root mean square error $\RMSE_X\approx1$\%). Nonetheless, the relatively high observed dispersion ($\rm \approx3-12\%$) suggests that the absorbance model described in this study still has its limitations. Unknown non-linear effects might be at play, opening the door for potential improvements of the absorbance model in future studies.

\begin{figure}[h!]
    \centering
    \includegraphics[scale=0.5]{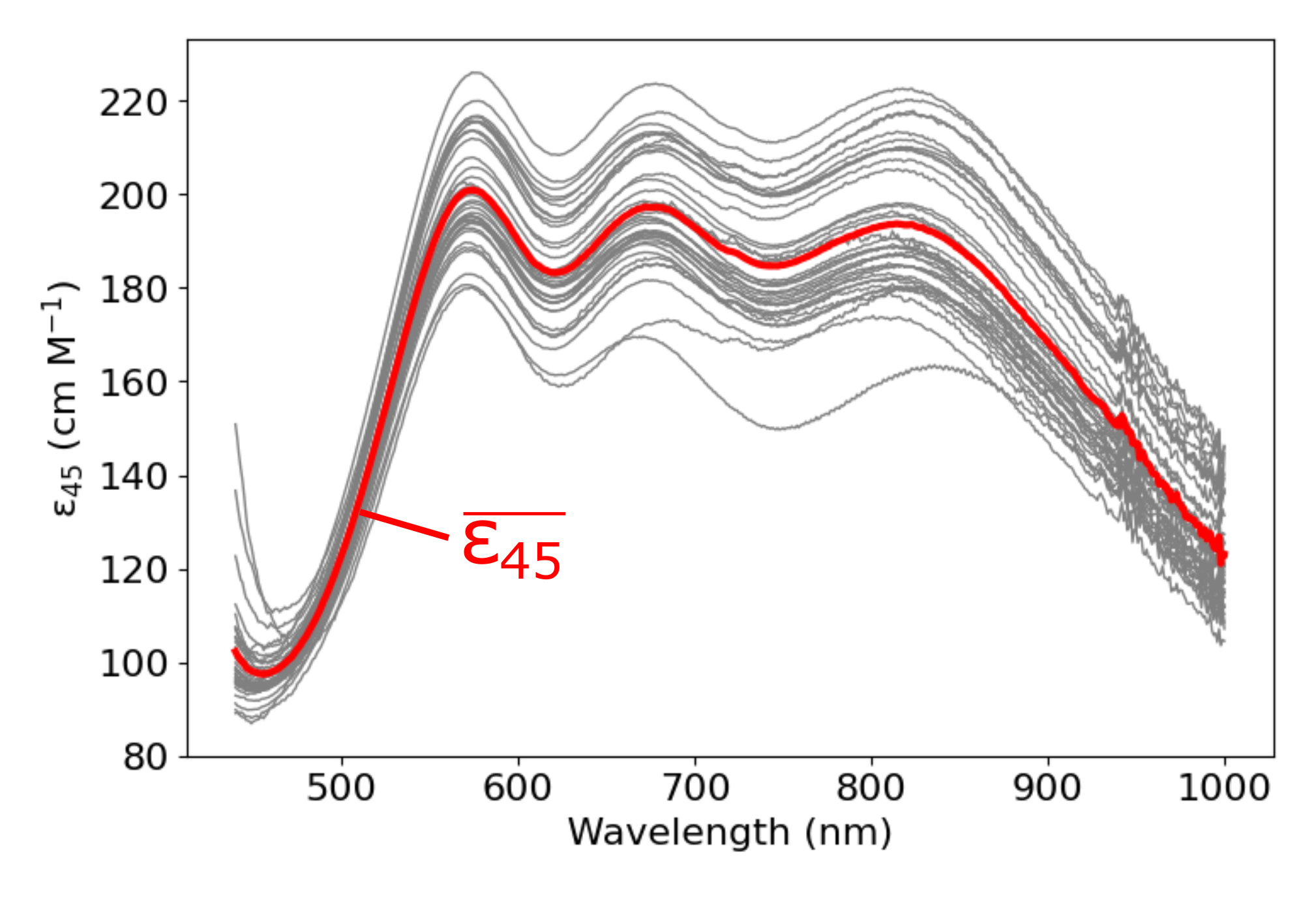}
    \caption{Molar absorptivity of the complex $\epsilon_{45}$ computed using Equation~(27) for each calibration sample of the positive electrolyte (gray lines), and average absorptivity (red line) used as reference for the calibration.}
\end{figure}

\section{Parametric study for the spectral deconvolution of the positive electrolyte}

\begin{figure}[b!]
    \centering
    \includegraphics[scale=0.5]{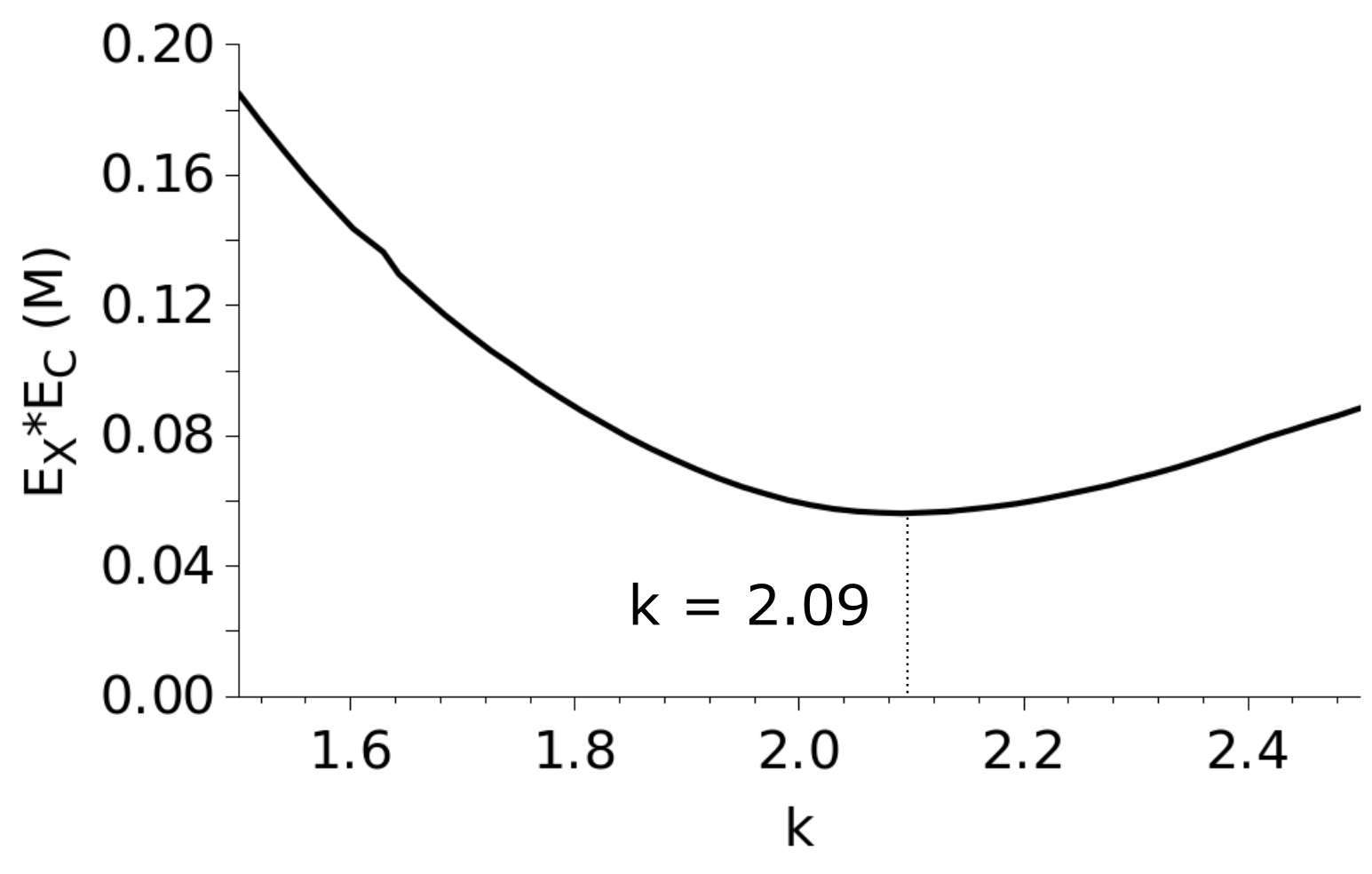}
    \caption{Product of $E_{X_4}$ and $E_C$ as a function of $k$ for the spectral deconvolution of the positive electrolyte. A clear minimum is reached for $k=2.09$.}
\end{figure}

This section presents a parametric study aimed at determining the optimal value of the power-law exponent $k$ that minimizes the overall experimental error in the spectral deconvolution of the positive electrolyte presented in Section 3.3.2 of the main text. Figure S4 illustrates the result of the study by plotting the product of the RMSE errors, $E_{X{_4}}$ and $E_C$, as a function of the factor $k$ involved in Equations (2), (27), and (28). Notably, the figure exhibits a clear minimum at $k=2.09$, indicating that this value represents the optimal choice for the deconvolution process. In fact, this is the value used in the results presented in Figure 8 and Table 8. It is worth noting that $2.09$ is consistent with the value $k = 1.88$ indicated in Figure~1g, computed using a single wavelength of 440 nm.